\documentclass[%
 reprint,
superscriptaddress,
nofootinbib,
 amsmath,amssymb,
 aps,
prd,
]{revtex4-2}

\usepackage{dcolumn}

\usepackage{graphicx} 
\usepackage{amsmath}
\usepackage{amssymb}
\usepackage[bottom]{footmisc}
\usepackage{braket}
\usepackage{slashed}
\usepackage{verbatim}
\usepackage{adjustbox}
\usepackage{enumitem} 
\usepackage[dvipsnames]{xcolor}
\usepackage{mathtools}

\usepackage{graphicx}
\usepackage{xcolor}
\usepackage{rotating}
\usepackage{bm,amsmath,amssymb}
\usepackage[utf8]{inputenc}
\usepackage{comment}
\usepackage{footmisc}
\usepackage{diagbox}
\usepackage{amsmath}
\usepackage{cancel}
\usepackage{physics}
\usepackage{dcolumn}
\usepackage{bm}

\usepackage[colorlinks=true, urlcolor=blue, linkcolor=blue, citecolor=red, hyperindex=true, linktocpage=true]{hyperref}

\usepackage{tikz}
\usetikzlibrary{matrix}
\usepackage{filecontents}

\usetikzlibrary{shapes.misc}
\tikzset{cross/.style={cross out, draw=black, minimum size=2*(#1-\pgflinewidth), inner sep=0pt, outer sep=0pt},
cross/.default={1pt}}

\def\CO{\mathcal{O}}

\newcommand{\pars}[1]{\left(#1\right)}
\newcommand{\spars}[1]{\left[#1\right]}

\newcommand{\dbar}{{\mathchar'26\mkern-12mu d}}
\newcommand{\deltabar}{{\mathchar'26\mkern-9mu \delta}}
\newcommand{\deps}{\delta_{(\eta)}}
\newcommand{\smgt}{ {\scalebox{.5}{$>$}} }
\newcommand{\smls}{ {\scalebox{.5}{$<$}} }
\newcommand{\smhash}{\vcenter{\hbox{\scalebox{.5}{\#}}}}

\begin{document}

\title{Wilsonian renormalisation group and thermal field theory \\in the Schwinger-Keldysh closed-time-path formalism}
\author{Giorgio Frangi}
\affiliation{
	Higgs Centre for Theoretical Physics, The University of Edinburgh, \\
	Peter Guthrie Tait Road, King's Buildings, Edinburgh EH9 3FD, Scotland
}
\author{Sa\v{s}o Grozdanov}
\affiliation{
	Higgs Centre for Theoretical Physics, The University of Edinburgh, \\
	Peter Guthrie Tait Road, King's Buildings, Edinburgh EH9 3FD, Scotland
}
\affiliation{
	Faculty of Mathematics and Physics, University of Ljubljana, \\
	Jadranska Ulica 19, SI-1000 Ljubljana, Slovenia
}

\begin{abstract}
We study perturbative Wilsonian renormalisation group (RG) for the scalar $\phi^4$ theory at finite temperature to one loop order in the Schwinger-Keldysh closed-time-path (CTP) formalism. By explicitly integrating out the UV modes, we show how effective interactions coupling the two branches of the CTP contour arise. This provides a microscopic derivation of the type of effective CTP actions used in the literature. While the full effective theory has no critical points, we instead proceed to study critical properties of the RG flow in a reduced space of complex couplings that characterise certain time-ordered correlators and show the existence of two novel fixed points in $d=4$ spacetime dimensions. We relate one to the Wilson-Fisher fixed point known to exist only in $4-\epsilon$ dimensions, and the other to the standard Gaussian fixed point.
\end{abstract}

\maketitle

\tableofcontents

\section{Introduction}
Quantum field theory (QFT) provides us with a description of nature at a fundamental, microscopic level. Such theories therefore typically govern physical systems at very small length scales (in the ultraviolet (UV) regime). By probing the same systems at large distance scales, or at low energies (the infrared (IR) limit), we observe a vast breadth of collective states of matter typically described by effective field theories (EFTs). The main theoretical procedure to elucidate how such IR EFTs arise from a UV QFT is known as the Wilsonian renormalisation group (RG). In Wilsonian RG, one integrates the UV degrees of freedom from the partition function and absorbs this `information' into couplings and also new operators of the theory. This results in an EFT for the IR degrees of freedom, which correctly captures the low-energy behaviour of the system. In many situations, the IR EFT contains only a finite subset of all possible quantum operators, making it simple. For a description of this procedure, see e.g.~Ref.~\cite{polchinskinotes}.

Very often, the Wilsonian RG procedure is performed at zero temperature ($T = 0$), starting from a functional integral defined between an initial ($t_i\to - \infty$) and final ($t_f\to\infty$) time, through which it is possible to compute the in-out scattering processes between asymptotically free vacua. Nevertheless, many physically relevant questions in condensed matter physics, cosmology and in particle physics (although less commonly), are of a different type: instead of computing scattering amplitudes or decay probabilities, we may be interested in the time evolution of an operator $\CO(t)$ given some initial state, for example, the thermal density matrix $\hat\rho \sim e^{-\beta \hat H}$, where $\beta = 1/T$ and $\hat H$ is the Hamiltonian of the system. In such cases, it is often necessary or at least convenient to use the real-time Schwinger-Keldysh formulation of QFT, also called the closed-time-path (CTP) or `in-in' formalism. As we discuss below this approach requires the degrees of freedom to time-evolve on a complex time contour, which can be interpreted as the doubling of the degrees of freedom. For an introduction to these techniques, see e.g.~Refs.~\cite{LandvWeert,Das:ftft,calzetta2008nonequilibrium,Kamenev_2011,cgl_lectures}.

In this work, we apply the Wilsonian RG analysis to a theory defined on a complex closed time path, and investigate the novel features of the resulting EFT.\footnote{For a very recent related paper, see Ref.~\cite{Green:2024cmx}.} Here, we do this in perhaps the simplest interacting QFT: the massive scalar $\phi^4$ theory with the bare Lagrangian density
\begin{equation}
    \label{eq:lagr_phi4}
    \mathcal L = \frac{1}{2} \partial_\mu \phi \partial^\mu \phi - \frac{1}{2} m^2 \phi^2 - g \phi^4.
\end{equation}
One goal is to establish the framework for future investigations of time-dependent processes in (non-equilibrium) EFTs. Moreover, this analysis is also motivated by past works on EFTs of dissipative hydrodynamics \cite{sasojanos1,Montenegro:2016gjq,Crossley_eft,Glorioso:IIthermo,cgl_part2,cgl_lectures,HaehlRang_a,HaehlRang_b,HaehlRang_sigma,Chen-Lin:2018kfl,Jensen:2017kzi,Jensen:2018hse}.\footnote{For other recent applications, see also Refs.~\cite{Vardhan:short,Vardhan:long,BlakeLiu1,BlakeLiu2,GaoLiu_nonmaximal,Winer:2020gdp,Jain:2020zhu,Abbasi:2024pwz,Montenegro:2017rbu,Akyuz:2023lsm,Baggioli:2023tlc,Hongo:2024brb,Armas:2024iuy,Jain:2023obu,Grozdanov:2024fle}.} In particular, we would like to better understand how dissipative (or, in some cases, open \cite{Nagy:2020bal,Kim:2023gaa}) EFTs arise from a closed, unitary QFT. A Wilsonian EFT derived on the CTP contour is expected to include a more complicated structure of relevant and marginal operators coupling the two time axes, sometimes referred to as the `influence functional' (see Ref.~\cite{sasojanos1}). Here, we show explicitly how this occurs and investigate some of the consequences of this enlarged space of new complex coupling constants. 

This process involves several additional complications as compared to the standard Wilsonian RG. Some of them are standard in CTP calculations and pertain to the extended time contour. Others, such as the appearance of pinching-pole singularities, are due to the fact that we are studying a QFT at finite temperature (see e.g.~Refs.~\cite{Jeon:spectral,jeon:long,Wang:2002nba,AMYpaper}). We address the latter issues by introducing a new prescription that regulates the pinching-pole singularities, which can in turn be motivated by physical requirements on the spectral density of the theory. Within this framework, renormalisation has a twofold effect: it gives an imaginary part to the real bare couplings $m$ and $g$, and also creates a new interaction between the two time branches of a single real time fold, which is multiplied by a purely imaginary coupling we call $g_\times$. In particular, the (dissipative) effective CTP action computed to one loop has the following unitarity-imposed structure:
\begin{align}\label{EFT_fin}
       &S_{\rm eff}[\phi_1,\phi_2] = \int d^dx \, \bigg[ \frac{1}{2} \partial_\mu \phi_1 \partial^\mu \phi_1 - \frac{1}{2}m^2(\zeta) \phi_1^2 \nonumber\\
       & - \frac{1}{2} \partial_\mu \phi_2 \partial^\mu \phi_2 + \frac{1}{2}m^2(\zeta) \phi_2^2 - g(\zeta) \phi^4_1 + \bar g(\zeta) \phi^4_2 \nonumber\\
       & + g_\times(\zeta) \phi^2_1 \phi_2^2 \bigg],
\end{align}
where $\zeta$ corresponds to the running RG scale, $\bar g$ is the complex conjugate of $g$ and $g_\times$ is imaginary. These predictions, discussed in Ref.~\cite{sasojanos1}, are in line with the expectation that the EFT should encode dissipation, and provide a microscopic foundation for the EFT framework mentioned above.

Equipped with explicit perturbative expressions for the running couplings, we then study the critical properties of the renormalisation flow in an enhanced complex space of $m$, $g$ and $g_\times$. While the choice of the UV theory \eqref{eq:lagr_phi4}---which effectively constrains the initial conditions of the RG flow---prevents us from finding interacting fixed points in the full space of couplings, restricting our analysis to a particular subclass of time-ordered correlators does in fact allow us to define a smaller coupling constant space (which we name `reduced') in which fixed points do exist. Surprisingly, this mathematical structure leads to a flow diagram reminiscent of a system with a Gaussian and a Wilson-Fisher fixed point exactly at the critical dimension, which can be continuously deformed to match the standard RG story of the two fixed points.

The structure of the paper is as follows. In Section~\ref{sec:spd_fr}, we discuss the setup and derive the relevant Feynman rules. Moreover, we discuss their relation with the spectral density. In Section~\ref{sec:wil_ren}, we perform the Wilsonian RG procedure and discuss the details of the CTP EFT. Then, in Section~\ref{sec:beta_complex}, we define the reduced space of complex couplings, which appear in correlation functions defined on a single leg of the CTP contour, and investigate the critical properties of such RG flows. We conclude with Section~\ref{sec:conc}. A few appendices have also been added at the end that spell out various technical details.

\section{The setup: the microscopic QFT}
\label{sec:spd_fr}

In this paper, we study the QFT of a single massive real scalar field at finite temperature $T = 1/\beta$ with the $\phi^4$ interaction introduced in Eq.~\eqref{eq:lagr_phi4}. In that expression, $m$ is the bare mass and $g$ the bare coupling constant. Adopting standard terminology from the RG literature, we at times collectively refer to $m$ and $g$ as `couplings'. We will also refer to the quadratic part of \eqref{eq:lagr_phi4} as $\mathcal L_\text{free}$, and the quartic interaction as $\mathcal L_\text{int}$.

\begin{figure}[t!]
    \centering
    \includegraphics[width=\linewidth]{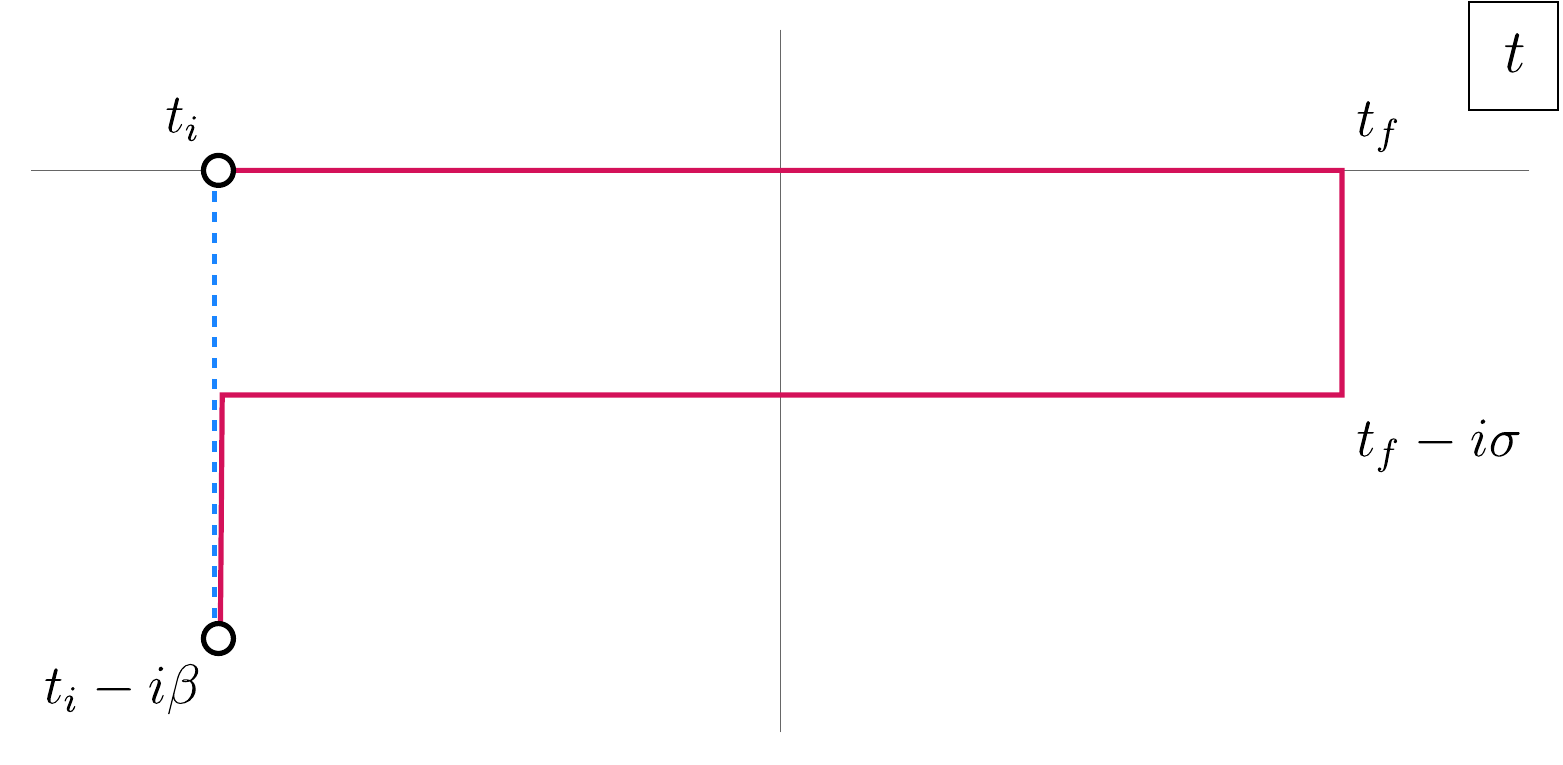}
    \caption{Two possible choices for time contours connecting some initial time $t_i$ and $t_i-i\beta$: Matsubara (blue, dashed), and a CTP contour (magenta). The latter choice, which allows for direct calculations of real-time observables, is really a family of choices labeled by $\sigma \in (0,\beta)$. The choice of $\sigma = \beta/2$ corresponds to the so-called thermofield double state (see e.g.~\cite{Das:ftft}).}
    \label{fig:time_contour}
\end{figure}

The CTP formulation of thermal QFT consists of placing the theory on a generic complex time path $\gamma$, starting at some initial time $t_i$ and ending at $t_f = t_i - i \beta$. Many choices for $\gamma$ are possible, see Figure~\ref{fig:time_contour} for a few examples. While the choice of the contour influences many features of the theory, e.g., its Feynman rules \cite{LandvWeert}, certain properties and, certainly, all physically measurable results of the calculations must be independent of the choice. 

One of those, of crucial importance to our analysis, is the relation between the spectral function of the theory and its propagators. We illustrate it here. Consider the generating functional of the free theory $\mathcal L_\text{free}$, defined as the action-weighted functional integral over field configurations $\phi$ with boundary conditions specified at $t_i$ and $t_i - i \beta$, which we impose to be periodic. By further assuming the Kubo-Martin-Schwinger (KMS) relation, the free generating functional can then be written as 
\begin{equation}
	\label{eq:free_PI_c}
	Z_{0}[j] = \exp \spars{-\frac{i}{2} \int dt dt' \, j(t) D_0^{(c)}(t-t') j(t')},
\end{equation}
where the time variables $t$ and  $t'$ run along the complex contour $\gamma$ and the integration on spatial volume is implied. The superscript $(c)$ denotes ordering on $\gamma$, and $j$ is a source for the $\phi$ field defined on it. Importantly, the free Green's function $i D_0^{(c)}$ is related to the free spectral density $\rho_{0}$ by:
\begin{equation}
	\label{eq:green_spectral}
	\begin{aligned}
		i D_0^{(c)}(t-t') &= \int \frac{d^d k}{(2\pi)^d} e^{- i k (x - x')} \rho_0(k)  \\
        &\times \left[ \theta_{c}(t-t') + n(k_0)  \right],
	\end{aligned}
\end{equation}
where $\theta_c$ is the Heaviside step function defined on $\gamma$, $n(k_0) = (e^{\beta k_0}-1)^{-1}$ is the Bose-Einstein distribution and
\begin{equation}
	\label{eq:free_spectral}
	\rho_0(k) = \lim_{\eta\to0} 2 \pi \,\text{sgn}(k_0) \deps(k^2 - m^2).
\end{equation}
Here, $\deps$ is a regularised delta function, which reduces to the Dirac delta function $\delta$ in the $\eta\to0$ limit:
\begin{equation}
    \label{eq:reg_delta}
    \deps(x) = \frac{1}{2\pi i} \pars{\frac{1}{x- i \eta} - \frac{1}{x + i \eta}} = \frac{1}{\pi} \frac{\eta}{x^2 + \eta^2}.
\end{equation}
The convenience of writing the spectral function as the limit of a regularised representation of the on-shell $\delta$ will become clear in the following. Crucially, by virtue of the KMS relation, Eq.~\eqref{eq:green_spectral} holds for interacting fields as well, provided that the free propagator and the spectral function are substituted with the interacting ones. As $\rho$ does not depend on the contour $\gamma$, this statement implies that all propagators can be derived from a single complex function (the spectral function) and the choice of $\gamma$ \cite{LandvWeert}.

We choose to adopt the magenta CTP contour in Figure~\ref{fig:time_contour}. Each of the time arguments in \eqref{eq:free_PI_c} can take values in any of the four segments $I_i$ of $\gamma$. By defining fields with support on one interval only:
\begin{equation}
    \phi_i(t,\mathbf x) = 
    \begin{cases}
        \phi(t,\mathbf x) & t \in I_i \\
        0 & \text{otherwise}
    \end{cases}
\end{equation}
one could expect $iD_0^{(c)}$ to be a rank-$4$ matrix. However, for real-time calculations, there is no need to keep the full structure. Assuming the free spectral function $\rho_0$ to be an ordinary function, and requiring the source $j$ to vanish in the distant past and future, the Riemann-Lebesgue lemma ensures that vertical and horizontal branches factorise in \eqref{eq:free_PI_c}. For the purpose of calculating real-time correlators, then, the former can be integrated out, leading to a simpler rank-$2$ propagator matrix structure. This gives rise to the familiar doubling of the degrees of freedom in thermal field theory:
\begin{equation}
    \label{eq:doubled_fields}
    \phi_1(t,\mathbf x) = \phi(t,\mathbf x), \quad \phi_2(t,\mathbf x) = \phi(t-i\sigma,\mathbf x),
\end{equation}
where $\sigma \in (0,\beta)$ is an arbitrary real parameter describing the distance between the two horizontal branches. One can also break analogously into two the contour-dependent source $j$. It is then convenient to rewrite the Lagrangian as in terms of a single real-time $t$:
\begin{equation}
	\label{eq:lagr_double}
	\mathcal L_{\rm CTP} = \mathcal L_1 - \mathcal L_2,
\end{equation}
where each $\mathcal L_i$ is obtained by substituting $\phi \to \phi_i$ in \eqref{eq:lagr_phi4}, and the minus sign takes into account of the different time orientation on the two branches.

In practice, the requirement that $\rho_0$ be an ordinary function---rather than a distribution---is fulfilled by taking the limit $\eta\to0$ in \eqref{eq:free_spectral} at the end of any calculation. Physically, this corresponds to giving $\phi$ a finite lifetime (of order $\eta^{-1}$) before taking it to be infinite at the end. We will work with the Feynman rules for finite $\eta$, which follow from rewriting \eqref{eq:free_PI_c} in terms of the fields \eqref{eq:doubled_fields}:
\begin{equation}
	\label{eq:free_PI_stf}
	Z_{0}[j] = \exp \left\{-\frac{i}{2} \int dt dt' \, j_r(t) D_0^{(rs)}(t-t') j_s(t') \right\},
\end{equation}
where $r$ and $s$ run over $\{1,2\}$. Explicitly,
\begin{equation}
    \label{eq:freeKG_mom}
    \begin{aligned}
        & i D_0^{(11)}(k) = \frac{i}{k^2-m^2 + i\eta} + 2 \pi n (|k_0|)\deps(k^2 - m^2) , \\
        & i D_0^{(22)}(k) = \frac{-i}{k^2-m^2 - i\eta} + 2 \pi n (|k_0|)\deps(k^2 - m^2), \\
        & i D_0^{(12)}(k) = 2 e^{\sigma k_0} \pi \left[ \theta(-k_0) + n (|k_0|) \right]\deps(k^2 - m^2), \\
        & i D_0^{(21)}(k) = 2 e^{- \sigma k_0} i \pi \left[ \theta(k_0) + n (|k_0|) \right] \deps(k^2 - m^2).
    \end{aligned}
\end{equation}
In what is to follow, we set $\sigma = \beta/2$ (the thermofield double choice), which makes the propagator symmetric under $r\leftrightarrow s$:
\begin{equation}
	\label{eq:freeKG_tfd}
        i D_0^\text{off-diag}(k) = 2 \pi e^{\beta|k_0|/2} n (|k_0|) \deps(k^2 - m^2).
\end{equation}

Next, to account for the quartic interaction, we write the generating functional as
\begin{equation}
    \label{eq:full_gf}
     Z[j] = \text{exp} \left\{ i \int d t \, \spars{ \mathcal L_\text{int} \pars{\frac{\delta}{i \delta j_1}} - \mathcal L_\text{int} \pars{\frac{\delta}{i \delta j_2}} } \right\} Z_0[j],
\end{equation}
where $Z_0[j]$ is given by \eqref{eq:free_PI_stf}, and $\mathcal L_\text{int}$ by \eqref{eq:lagr_phi4}. This gives us the Feynman rules for the full theory. We will denote different propagator lines as
\begin{gather}
      \vcenter{\hbox{\begin{tikzpicture}[scale=1]
        \node (1) at (0,0) {1};
        \node (2) at (2,0) {1};
        \draw
            (1) edge[-] (2);
        \end{tikzpicture}}} = i D_0^{(11)}(k)  \text{ in \eqref{eq:freeKG_mom}}, \\
       \vcenter{\hbox{\begin{tikzpicture}[scale=1]
        \node (1) at (0,0) {2};
        \node (2) at (2,0) {2};
        \draw
            (1) edge[-,dashed] (2);
        \end{tikzpicture}}} =i D_0^{(22)}(k)  \text{ in \eqref{eq:freeKG_mom}}, \\
        \begin{aligned}
        		\vcenter{\hbox{\begin{tikzpicture}[scale=1]
        		\node (1) at (0,0) {1};
        		\node[inner sep=0pt] (mid) at (1,0) {};
       		\node (2) at (2,0) {2};
        		\draw (1) edge[-] (mid);
        		\draw (mid) edge[-,dashed] (2);
        		\end{tikzpicture}}} = 
        		\vcenter{\hbox{\begin{tikzpicture}[scale=1]
        		\node (1) at (0,0) {2};
        		\node[inner sep=0pt] (mid) at (1,0) {};
        		\node (2) at (2,0) {1};
        		\draw
            		(1) edge[-,dashed] (mid)
            		(mid) edge[-] (2);
        		\end{tikzpicture}}}&  = \\
        = i D_0^\text{off-diag}&(k)  \text{ in \eqref{eq:freeKG_tfd}},
        \end{aligned}
\end{gather}
and the vertices as
\begin{equation}
    \vcenter{\hbox{\begin{tikzpicture}
        \node (1) at (0,0) {1};
        \node (2) at (1.4,0) {1};
        \node (3) at (1.4,1.4) {1};
        \node (4) at (0,1.4) {1};
        \draw
            (1) edge[-] (3)
            (2) edge[-] (4);
        \node[circle,fill=black,inner sep=0pt,minimum size=4pt] (v) at (.7,.7) {};
    \end{tikzpicture}}} = - i 4! g, \quad 
    \vcenter{\hbox{\begin{tikzpicture}
        \node (1) at (0,0) {2};
        \node (2) at (1.4,0) {2};
        \node (3) at (1.4,1.4) {2};
        \node (4) at (0,1.4) {2};
        \draw
            (1) edge[-,dashed] (3)
            (2) edge[-,dashed] (4);
        \node[circle,draw=black,fill=white,inner sep=0pt,minimum size=4pt] (v) at (.7,.7) {};
    \end{tikzpicture}}} = i 4! g.
\end{equation}
Finally, we note that while it is sometimes convenient to work in the Keldysh basis with retarded and advanced fields (see Ref.~\cite{Kamenev_2011}), we find that the `1-2' basis is nevertheless best suited for present purposes.

\section{Wilsonian renormalisation}
\label{sec:wil_ren}

We now turn our attention to the study of thermal IR properties of $\phi^4$ theory \eqref{eq:lagr_phi4} on the CTP contour, which allows for analysis of real-time observables. We will do this by using a procedure analogous to the textbook Wilsonian RG with a hard UV cutoff. We will also work in a general number of spacetime dimensions $d$.

More precisely, we assume that that the original bare theory has a finite UV cutoff, which restricts the measure of the generating functional to:
\begin{equation}
    D \phi(x) = \prod_{k < \Lambda} d \phi_1(k) d \phi_2(k).
\end{equation}
The renormalisation procedure then consists of the following steps:
\begin{enumerate}
	\item start by selecting a new, lower cutoff $\Lambda' < \Lambda$, which divides the Fourier modes of $\phi$ into slow ($\phi_\smls, \, k<\Lambda'$) and fast ($\phi_\smgt, \, \Lambda'<k<\Lambda$);
	\item integrate out the fast modes to obtain an effective theory $S_{\rm eff}[\phi_\smls]$ of the slow modes, defined up to a new cutoff $\Lambda'$;
	\item rescale momenta and distances to find a theory valid up to $\Lambda$.
\end{enumerate}
In terms of the path integral, the effective action is then given by
\begin{equation}
	\label{eq:eff_act_def}
		e^{i S_{\rm eff}[\phi_\smls]} \equiv e^{i S_0[\phi_\smls]} \int D\phi_\smgt e^{i S_0[\phi_\smgt]} e^{i S_\text{int}[\phi_\smls,\phi_\smgt]}.
\end{equation}
Notice that the free part of the action, once written in momentum space, factorises into a free action for fast and slow modes separately. This is a consequence of it being quadratic, which also allows us to consider the path integral over fast modes as an average over a Gaussian ensemble. A few lines of algebra lead to:
\begin{equation}
	i S_0[\phi_\smgt] = -\frac{1}{2} \int \dbar k \,\dbar k'\phi_{(r)}^\smgt(k) G_{(rs)}^{-1}(k,k') \phi_{(s)}^\smgt(k'),
\end{equation}
where $\dbar k = d^dk/(2\pi)^d$ and:
\begin{equation}
	\label{eq:green_Grs}
	G^{(rs)}(k,k') = i D_{0}^{(rs)}(k) (2\pi)^d \delta(k+k').
\end{equation}
This allows us to write the path integral in \eqref{eq:eff_act_def} as
\begin{equation}
	\int D\phi_\smgt \, e^{i S_0[\phi_\smgt]} e^{i S_\text{int}[\phi_\smls,\phi_\smgt]} = \langle  e^{i S_\text{int}[\phi_\smls,\phi_\smgt]} \rangle_\smgt, 
\end{equation}
so that
\begin{equation}
	\label{eq:eff_act_impr}
		e^{i S_{\rm eff}[\phi_\smls]} = e^{i S_0[\phi_\smls]}  \langle  e^{i S_\text{int}[\phi_\smls,\phi_\smgt]} \rangle_\smgt.
\end{equation}
Treating the interacting part as a perturbation, which we take to be controlled by a dimensionless coupling (such as $\Lambda^{4-d} g$), and taking the logarithm of the expression, we find
\begin{equation}
	\label{eq:diag_exp_eff}
	\begin{aligned}
		S_{\rm eff}[\phi_\smls] \simeq S_\text{0}[\phi_\smls] + \langle S_\text{int}[\phi_\smls,\phi_\smgt] \rangle_\smgt & + \\ + \frac{i}{2} ( \langle S_\text{int}[\phi_\smls,\phi_\smgt]^2 \rangle_\smgt - \langle & S_\text{int}[\phi_\smls,\phi_\smgt] \rangle_\smgt^2 ).
	\end{aligned}
\end{equation}
Finally, what remains to be done is to take the averages over the fast modes, and read off the corrections to the couplings. This can be done either by using Wick's theorem or by computing Feynman diagrams. Because \eqref{eq:diag_exp_eff} takes the form of a cumulant expansion, only connected diagrams need to be considered.

\subsection{Mass renormalisation}

The interacting action can be rewritten in terms of Fourier modes as
\begin{equation}
    \label{eq:interacting_action}
    S_\text{int}[\phi_1,\phi_2] = \sum_{s=1,2}(-1)^s g \int \prod_{j=1}^4 \dbar k_j \phi_s(k_j)\, \deltabar\pars{\textstyle\sum\nolimits_i k_i},
\end{equation}
where we defined the symbol $\deltabar(k) = (2\pi)^d \delta(k)$ so that
\begin{equation}
	\int \dbar k \deltabar (k) = 1.
\end{equation}
The form of Eq.~\eqref{eq:interacting_action} implies that in calculating the linear correction to the effective action, fields defined on the two CTP branches do not mix with each other; thus, we focus on corrections involving $\phi_1$ only (the ones involving $\phi_2$ will differ by an overall minus sign and give no additional contribution to the mass renormalisation). Now, each of the four modes in Eq.~\eqref{eq:interacting_action} can be either fast or slow, giving in total $2^4=16$ combinations. By relabeling momenta, and suppressing the index 1 (all fields are of that type), we find that
\begin{equation}
    \langle S_\text{int}[\phi_\smls,\phi_\smgt] \rangle_\smgt= - g \int \prod_{j=1}^4 \dbar k_j \langle \mathcal C(k_j) \rangle_\smgt \, \deltabar\pars{\textstyle\sum\nolimits_i k_i},
\end{equation}
where we defined the function $\mathcal C(k_j)$ as follows:
\begin{align}
        &\mathcal C(k_j) = \phi_{k_1}^\smls \phi_{k_2}^\smls \phi_{k_3}^\smls \phi_{k_4}^\smls +  4 \phi_{k_1}^\smls \phi_{k_2}^\smls \phi_{k_3}^\smls \phi_{k_4}^\smgt  \\ &+  6 \phi_{k_1}^\smls \phi_{k_2}^\smls \phi_{k_3}^\smgt \phi_{k_4}^\smgt + 4 \phi_{k_1}^\smls \phi_{k_2}^\smgt \phi_{k_3}^\smgt \phi_{k_4}^\smgt  +  \phi_{k_1}^\smgt  \phi_{k_2}^\smgt  \phi_{k_3}^\smgt \phi_{k_4}^\smgt. \nonumber
\end{align}
Taking the average over fast modes then gives to leading order the expression
\begin{equation}
   - 6 g \int \dbar k_1 \,\dbar k_2 \,  \phi_{k_1}^\smls \phi_{k_2}^\smls \int \dbar k_3 \,\dbar k_4 \,  \langle \phi_{k_3}^\smgt \phi_{k_4}^\smgt  \rangle_\smgt \deltabar \pars{\textstyle\sum\nolimits_i k_i}.
\end{equation}
The two-point function that appears in this  expression is $G^{(11)}$ in Eq.~\eqref{eq:green_Grs}. Hence, we obtain
\begin{equation}
	\delta S[\phi_\smls] = - \frac{1}{2} \delta m^2 \int \dbar k_1 \,\dbar k_2 \, \phi_{k_1} \phi_{k_2} \deltabar(k_1 + k_2), 
\end{equation}
where the correction to the bare mass is given by the following one-loop integral:
\begin{equation}
	\label{eq:m2_corr}
	\delta m^2 = 12 g \int_\smgt \dbar p \, i D^{(11)}(p) \sim  \vcenter{\hbox{\begin{tikzpicture}
            	\node[circle,draw=black,fill=black,inner sep=0pt,minimum size=4pt] (v) at (0,0) {};
            	\draw (v) arc [radius=.4, start angle=-90, end angle= 270];
        		\end{tikzpicture}}}.
\end{equation}


To obtain an explicit expression for $\delta m^2$, we must now choose the cutoff scheme more precisely. At $T=0$, hard cutoffs such as the one we adopt are typically implemented by Wick-rotating the theory to Euclidean signature and integrating out Fourier modes in a $d$-dimensional ball of radius $\Lambda$. The results are then analytically continued back to Lorentzian signature. For a recent discussion of explicit Lorentzian cutoffs, see Refs.~\cite{Polonyi:2018ykh,Steib:2019xrv}.
In a thermal state with $T \neq0$, Lorentz invariance is explicitly broken (see e.g.~Ref.~\cite{Ojima:nice} for a detailed discussion). If the cutoff prescription is supposed to respect the symmetries of the theory and the state (here, spatial rotations), this means that we now have more freedom. We therefore propose to impose a cutoff on spatial momenta only, and leave the loop integrals over frequencies unbounded. In terms of the imaginary time quantisation, this would correspond to cutting the loop momenta off after having performed the infinite discrete sums over Matsubara frequencies. Similar procedures can be found in some earlier works, for example in \cite{ElmforsKobes_thbeta,Chen-Lin:2018kfl}.

Equipped with this prescription, we now calculate the mass correction \eqref{eq:m2_corr}:
\begin{equation}
	\label{eq:m_corr}
	m^2 \to m^2 + \frac{12 g \Omega_{d-2}}{(2\pi)^{d-1}} \int_{\Lambda'}^{\Lambda} dp \frac{p^{d-2}}{2 E_p} \coth\pars{\frac{\beta E_p}{2}},
\end{equation}
where $\Omega_{d}$ is the surface area of the $d-$dimensional sphere and
\begin{equation}
	E_p = \sqrt{p^2 + m^2}.
\end{equation}
It is easy to check that by taking $\Lambda'\to0$, the result matches the one-loop mass correction computed in \cite{Das:ftft}.

\subsection{Coupling constant renormalisation}

Next, we compute the leading-order (one-loop) quantum correction to the bare quartic coupling $g$. The relevant correction arises at second order in the interactions, and is encoded in the connected part with four slow modes (of any of the two types) at vanishing external momentum. There are three such contributions, respectively coming from
\begin{widetext}
\begin{equation}
	\label{eq:corr1}
	\mathcal I_1 \! = i 18 g^2 \!\! \int_\smls \dbar k_1 \, \dbar k_2\,  \dbar p_1 \, \dbar p_2 \, \phi_1^\smls(k_1) \phi_1^\smls(k_2) \phi_1^\smls(p_1) \phi_1^\smls(p_2) \!\int_\smgt \dbar k_3 \, \dbar k_4\,  \dbar p_3 \, \dbar p_4 \, \langle \phi_1^\smls(k_3) \phi_1^\smls(k_4) \phi_1^\smls(p_3) \phi_1^\smls(p_4) \rangle_\smgt \deltabar \pars{\textstyle\sum\nolimits_i k_i} \deltabar\pars{\textstyle\sum\nolimits_i p_i},
\end{equation}
\begin{equation}
	\label{eq:corr2}
	\mathcal I_\times\!  =  \!- i 36 g^2 \!\! \int_\smls \dbar k_1 \, \dbar k_2\,  \dbar p_1 \, \dbar p_2 \, \phi_1^\smls(k_1) \phi_1^\smls(k_2) \phi_2^\smls(p_1) \phi_2^\smls(p_2) \!\int_\smgt \dbar k_3 \, \dbar k_4\,  \dbar p_3 \, \dbar p_4 \, \langle \phi_1^\smls(k_3) \phi_1^\smls(k_4) \phi_2^\smls(p_3) \phi_2^\smls(p_4) \rangle_\smgt \deltabar \pars{\textstyle\sum\nolimits_i k_i} \deltabar\pars{\textstyle\sum\nolimits_i p_i},
\end{equation}
\begin{equation}
	\label{eq:corr3}
	\mathcal I_2 \!  = i 18 g^2 \!\! \int_\smls \dbar k_1 \, \dbar k_2\,  \dbar p_1 \, \dbar p_2 \, \phi_2^\smls(k_1) \phi_2^\smls(k_2) \phi_2^\smls(p_1) \phi_2^\smls(p_2) \! \int_\smgt \dbar k_3 \, \dbar k_4\,  \dbar p_3 \, \dbar p_4 \, \langle \phi_2^\smls(k_3) \phi_2^\smls(k_4) \phi_2^\smls(p_3) \phi_2^\smls(p_4) \rangle_\smgt \deltabar \pars{\textstyle\sum\nolimits_i k_i} \deltabar\pars{\textstyle\sum\nolimits_i p_i}.
\end{equation}
\end{widetext}
While $\mathcal I_1$ and $\mathcal I_2$ correspond to $O(g^2)$ corrections to $g$, $\mathcal I_\times$ gives rise to an entirely new RG-generated cross-coupling between the two CTP time axes. This effective coupling, which we call $g_\times$, is entirely absent in the original single-time-axis analysis. Its appearance is consistent with the proposal from Ref.~\cite{sasojanos1}. Interestingly, the terms $\phi_1\phi_2^3$ or $\phi_1^3\phi_2$ are not generated. At this perturbative order, this can be understood in terms of kinematic constraints. For example, the only diagram contributing to $\phi_1^3 \phi_2$ is:
\medskip
\begin{equation}
    \vcenter{\hbox{\begin{tikzpicture}
            \node[inner sep=0pt] (lu) at (-1.5,.5) {1};
            \node[inner sep=0pt] (ld) at (-1.5,-.5) {1};
            \node[inner sep=0pt] (lm) at (-1.7,0) {1};
            \node[circle,draw=black,fill=black,inner sep=0pt,minimum size=4pt] (lv) at (-1,0) {};
            \draw
                (lu) edge[-,dotted] (lv)
                (lm) edge[-,dotted] (lv)
                (ld) edge[-,dotted] (lv);
            \node[inner sep=0pt] (rd) at (1,-.5) {2};
            \node[inner sep=0pt] (rv) at (.5,0) {};
            \draw
                (rd) edge[-,dotted] (rv)
                (lv) edge[-] (-.25,0)
                (-.25,0) edge[-,dashed] (rv);
            \draw[dashed] 
                (.5,0) arc [radius=.3, start angle=-90, end angle= 270];
           \node[circle,draw=black,fill=white,inner sep=0pt,minimum size=4pt] at (.5,0) {};
        \end{tikzpicture}}},
\end{equation}
which vanishes as momentum conservation requires the momentum carried in the mixed internal line (a fast mode) to be the same as the one carried in the external type-2 leg (a slow mode). An analogous argument applies to $\phi_1 \phi_2^3$.

Using Wick's theorem and Eq.~\eqref{eq:green_Grs} to expand the connected 4-point functions in \eqref{eq:corr1}--\eqref{eq:corr3}, we find that
\begin{align}
   \mathcal I_1 = & \int \prod_{j=1}^4 \dbar p_j \phi_1(p_j)\, \delta \mathcal F_1(p) \deltabar\pars{\textstyle\sum\nolimits_i p_i}, \\
   \mathcal I_\times = & \int \prod_{j=1}^4 \dbar p_j \phi_{\left \lceil{j/2}\right \rceil }(p_j)\, \delta \mathcal F_\times(p) \deltabar\pars{\textstyle\sum\nolimits_i p_i}, \\
   \mathcal I_2 = & \int \prod_{j=1}^4 \dbar p_j \phi_2(p_j)\, \delta \mathcal F_2(p) \deltabar\pars{\textstyle\sum\nolimits_i p_i},
\end{align}
with $p = p_1 + p_2$, where $p_1$ and $p_2$ are the momenta of the two external modes attached to the same vertex, and the functions $\delta \mathcal F_i$ are defined as
\begin{align}
    \delta \mathcal F_1(p) = - i 36 g^2 \int \dbar k iD^{(11)}(k) iD^{(11)}(k-p), \\
    \delta \mathcal F_\times(p) = - i 72 g^2 \int \dbar k iD^{(12)}(k) iD^{(12)}(k-p), \\
    \delta \mathcal F_2(p) = i 36 g^2 \int \dbar k iD^{(22)}(k) iD^{(22)}(k-p).
\end{align}
To extract corrections to the quartic coupling, it is enough to take $p\to0$ rather than calculating the full loop integrals; as $p$ is the momentum of some external mode, an expansion in powers of $p^2$ creates effective interactions of the type $\phi^3\partial^{2k}\phi$, which are irrelevant in $d>2$. In terms of the couplings, the corrections \eqref{eq:corr1}--\eqref{eq:corr3} read:
\begin{align}
    \label{eq:coupl_corr1}
    g^{(1)} \to g - i 36 g^2 \int \dbar k \spars{iD^{(11)}(k)}^2, \\
    \label{eq:coupl_corrc}
    g_\times \to 0 - i 72 g^2 \int \dbar k \spars{iD^{(12)}(k)}^2, \\
    \label{eq:coupl_corr2}
    g^{(2)} \to g + i 36 g^2 \int \dbar k \spars{iD^{(22)}(k)}^2.
\end{align}
A corollary of this result is that the renormalised coupling constants determining the strength of the quartic interaction on each branch are different, as stressed by the notation. We will see below that this difference is necessary for the resulting EFT to be consistent with the microscopic unitary time evolution of the UV QFT.

Let us now analyse the corrections. Inserting the explicit expression for the propagators $iD^{(rs)}$ from Eqs.~\eqref{eq:freeKG_mom}--\eqref{eq:freeKG_tfd}, we run into a problem. This becomes apparent by isolating the frequency integrals from the loops; for example, we see that in \eqref{eq:coupl_corr1},
\begin{equation}
	\delta g^{(1)} = - i \frac{36 g^2 \Omega_{d-2}}{(2\pi)^{d-1}}\int dk \, k^{d-2} I_\omega^{(1)},
\end{equation}
where
\begin{align}
		&I_\omega^{(1)} = \int \frac{d\omega}{2\pi} \bigg[  \bigg( \frac{i}{\omega^2 - E_k^2 + i \eta} \bigg)^2 \\ 
        &+  \frac{i 4 \pi n(|\omega|) \deps(\omega^2 - E_k^2) }{\omega^2 - E_k^2 + i \eta}  + \big[2 \pi n(|\omega|) \deps(\omega^2 - E_k^2) \big]^2\bigg].\nonumber
\end{align}
The first of the two terms can be rewritten by using the following identity (see Ref.~\cite{NiemiSemenoff_nuclb} and Appendix~\ref{app:delta}):
\begin{equation}
    \label{eq:niemi}
	\frac{\deps(x^2 - \alpha^2)}{x^2 -\alpha^2 \pm i \eta} = - \frac{1}{2} \deps^{(1)}(x^2 - \alpha^2) \mp i \pi \left[\deps(x^2 -\alpha^2)\right]^2,
\end{equation}
where $\deps^{(1)}$ is the derivative of $\deps$. Thus, $I_\omega^{(1)}$ can be separated into a regular and a singular piece $I_\omega^{\rm sing}$:
\begin{equation}
	I_\omega^{(1)} = -i \frac{\beta E_k + \sinh{(\beta E_k)}}{8 E_k^3 \sinh^2{(\beta E_k /2 )}} + I_\omega^{\rm sing},
\end{equation}
where 
\begin{equation}
	\label{eq:ill_part}
	 I_\omega^{\rm sing} = \int \frac{d\omega}{2\pi}[2 \pi e^{\beta|\omega|/2} n(|\omega|) \deps(\omega^2 - E_k^2) \big]^2.
\end{equation}
Repeating the same procedure on Eqs.~\eqref{eq:coupl_corrc} and \eqref{eq:coupl_corr2},
\begin{align}
    \delta g_\times &= - i \frac{72 g^2 \Omega_{d-2}}{(2\pi)^{d-1}}\int dk \, k^{d-2} I_\omega^{(\times)}, \\
    \delta g^{(2)} &= i \frac{36 g^2 \Omega_{d-2}}{(2\pi)^{d-1}}\int dk \, k^{d-2} I_\omega^{(2)},
\end{align}
we then obtain the following results:
\begin{align}
    I_\omega^{(\times)} &=  I_\omega^{\rm sing}, \\
    I_\omega^{(2)} &= i \frac{\beta E_k + \sinh{(\beta E_k)}}{8 E_k^3 \sinh^2{(\beta E_k /2 )}} + I_\omega^{\rm sing}.
\end{align}

The fact that the same singular structure appears in every contribution can be understood as a corollary of a theorem that ensures that in perturbative calculations of two-point functions, superficially singular terms---i.e., products of $\delta$ functions---must sum to zero at every order in perturbation theory \cite{LandvWeert}. In the specific case of the $D^{(11)}$ propagator, the theorem ensures that
\begin{equation}
	   \vcenter{\hbox{\begin{tikzpicture}
            	\node[circle,draw=black,fill=black,inner sep=0pt,minimum size=4pt] (v) at (0,0) {};
		        \node[circle,draw=black,fill=black,inner sep=0pt,minimum size=4pt] (v2) at (0,.8) {};
            	\draw (v) arc [radius=.4, start angle=-90, end angle= 270];
                \draw (v2) arc [radius=.4, start angle=-90, end angle= 270];
        		\end{tikzpicture}}} + 
        \vcenter{\hbox{\begin{tikzpicture}
            	\node[circle,draw=black,fill=black,inner sep=0pt,minimum size=4pt] (v) at (0,0) {};
            	\draw (-.4,.4) arc [radius=.4, start angle=180, end angle= 360];
                \draw[dashed] (-.4,.4) arc [radius=.4, start angle=180, end angle= 0];
                \draw[dashed] (0,.8) arc [radius=.4, start angle=-90, end angle= 270];
                \node[circle,draw=black,fill=white,inner sep=0pt,minimum size=4pt] at (0,.8) {};
        		\end{tikzpicture}}} = \text{finite}.
\end{equation}
By direct inspection, one can check that 
\begin{equation}
    \label{eq:insp_eq}
	\vcenter{\hbox{\begin{tikzpicture}
            	\node[cross=3pt] (v) at (0,0) {};
            	\draw (v) arc [radius=.4, start angle=-90, end angle= 270];
        		\end{tikzpicture}}} =  \vcenter{\hbox{\begin{tikzpicture}
            	\draw[dashed] (0,0) arc [radius=.4, start angle=-90, end angle= 270];
		        \node[cross=3pt] (v) at (0,0) {};
        		\end{tikzpicture}}}\;,
\end{equation}
where the crosses indicate the vertices that have been divided out. This, in turn, implies that
\begin{equation}
    \label{eq:funny_div}
	   \vcenter{\hbox{\begin{tikzpicture}
            	\node[cross=3pt] (v) at (0,0) {};
		        \node[cross=3pt] (v2) at (0,.8) {};
            	\draw (v) arc [radius=.4, start angle=-90, end angle= 270];
        		\end{tikzpicture}}} - 
        \vcenter{\hbox{\begin{tikzpicture}
            	\node[cross=3pt] (v) at (0,0) {};
            	\draw (-.4,.4) arc [radius=.4, start angle=180, end angle= 360];
                \draw[dashed] (-.4,.4) arc [radius=.4, start angle=180, end angle= 0];
                \node[cross=3pt] at (0,.8) {};
        		\end{tikzpicture}}} = \text{finite},
\end{equation}
which means that, indeed, the singular terms of the above loop diagrams cancel. However, this theorem does not apply to the calculation of corrections to the coupling $g$, as these are encoded in the connected four-point function; consequently, pathological terms there do not cancel, so that Eq.~\eqref{eq:ill_part} needs to be regulated independently. We stress that this additional need for regulating divergences in $g$ does not change the results of the two-point functions. 

To make sense of Eq.~\eqref{eq:ill_part}, we calculate it at a small but finite value of the regulator $\eta$. Details are presented in Appendix~\ref{app:higher_PPs}. A direct application of Eq.~\eqref{eq:fint_eps} then gives the singular term
\begin{equation}
    \label{eq:ill_part2}
	 I_\omega^{\rm sing} = \frac{n(E_k)}{2 \pi E_k} \frac{1}{\eta} + O(\eta),
\end{equation}
which diverges as $\eta\to0$. The origin of this divergence can be traced back to the $p\to0$ limit of Eqs.~\eqref{eq:coupl_corr1}--\eqref{eq:coupl_corr2}, which causes simple poles of the integrand in the complex $k_0$ plane to collide. This is the well-known pinching-pole singularity of thermal field theories, which is responsible for the breakdown of perturbation theory in the IR regime of thermal correlators of conserved field bilinears (currents) \cite{jeon:long}. We draw inspiration from its physical interpretation to propose a regularisation scheme for \eqref{eq:ill_part}. Indeed, as we argued before, the limit $\eta\to0$ has the effect of making $\phi$ a stable particle with an infinite lifetime. In thermal field theory, however, interactions with the thermal bath give $\phi$ a finite spectral width. If the theory is weakly coupled, the resulting spectral function has the form
\begin{align}
	\label{eq:int_spectral}
	\rho(k) &= \frac{1}{E_k} \bigg[ \frac{\Gamma_k}{(k_0 - E_k)^2+\Gamma_k^2} -\frac{\Gamma_k}{(k_0 + E_k)^2+\Gamma_k^2} \bigg] \nonumber\\
    &\times \left( 1 + O(g^2)\right),
\end{align}
where the ($k$-dependent) thermal width $\Gamma_k$ is defined as
\begin{equation}
	\Gamma_k = - \frac{\text{Im}\pi(E_k,k)}{2 E_k} \sim O(g^2),
\end{equation}
and the perturbative order of the imaginary part of the boson self-energy is determined by the lowest perturbative order at which it appears (see, e.g., Ref.~\cite{Stanford_wkchaos} for a detailed discussion). This clarifies that the $\eta^{-1}$ divergence in \eqref{eq:ill_part2} is an artefact of only partially taking into account the thermal effects. A full treatment of this problem would require us to use the full renormalised perturbation theory at finite $T$ and using those results in the Wilsonian procedure. Due to the aforementioned breakdown of perturbation theory, this, in general, requires performing infinite resummations of Feynman diagrams, making it a technically difficult task. 

An alternative and significantly simpler pathway to, at least, a qualitative analysis, is offered by the fact that, regardless of the details, the spectral function has to satisfy Eq.~\eqref{eq:green_spectral}. Changing the spectral function to accommodate thermal effects can then correct the Feynman rules, and this correction can be used in the calculations. Using the identities presented in Appendix \ref{app:delta}, the regularised free spectral function can be written as
\begin{equation}
	\rho_{(\eta)}(k)  = \frac{1}{E_k} \left[\frac{\tilde\eta}{(k_0 - E_k)^2+\tilde\eta^2} - \frac{\tilde\eta}{(k_0 + E_k)^2+\tilde\eta^2} \right],
\end{equation}
where $\tilde\eta = \eta/2 E_k$. In order to take into account thermal effects, then, instead of taking the limit $\eta\to0$ at the end of the procedure, we require that
\begin{equation}
    \label{eq:hash_brown}
    \eta \to 2 E_k \kappa g^2,
\end{equation}
where we will leave $\kappa$ unspecified, and treat it throughout this work as a phenomenological constant that characterises the thermal width of $\phi$.
Using this prescription, we obtain a regularised version of $I_\omega^{\rm sing}$:
\begin{equation}
	I_\omega^{\rm sing} \to I_\omega^\text{reg} = \frac{1}{8 E_p^2 \kappa g^2 \sinh^2(\beta E_p/2)}.
\end{equation}
Substituting the regularised expression back into Eqs.~\eqref{eq:coupl_corr1}--\eqref{eq:coupl_corr2} then leads us to the following expressions for the single-time-leg couplings:
\begin{align}
	g^{(1)} &\to g - \frac{36 g^2 \Omega_{d-2}}{(2 \pi)^{d-1}} \int_{\Lambda'}^{\Lambda} dp \, \frac{p^{d-2}(\beta E_p+\sinh{\beta E_p})}{8 E_p^3 \sinh^2 (\beta E_p/2)} \nonumber \\ 
    &- i \frac{36 \Omega_{d-2}}{(2 \pi)^{d-1}} \int_{\Lambda'}^{\Lambda} dp \frac{p^{d-2}}{8 E_p^2 \kappa \sinh^2(\beta E_p/2)}, 	\label{eq:g_corr} \\
    g^{(2)} &\to g - \frac{36 g^2 \Omega_{d-2}}{(2 \pi)^{d-1}} \int_{\Lambda'}^{\Lambda} dp \, \frac{p^{d-2}(\beta E_p+\sinh{\beta E_p})}{8 E_p^3 \sinh^2 (\beta E_p/2)} \nonumber \\ 
    &+ i \frac{36 \Omega_{d-2}}{(2 \pi)^{d-1}} \int_{\Lambda'}^{\Lambda} dp \frac{p^{d-2}}{8 E_p^2 \kappa \sinh^2(\beta E_p/2)}, 	\label{eq:g2_corr}
\end{align}
with the new coupling between the two time axes $\phi_1^2\phi_2^2$ becoming
\begin{equation}
    \label{eq:gc_corr}
	g_\times = - i \frac{72 \Omega_{d-2}}{(2 \pi)^{d-1}} \int_{\Lambda'}^{\Lambda} dp \frac{p^{d-2}}{8 E_p^2 \kappa \sinh^2(\beta E_p/2)}.
\end{equation}
It is important to stress that due to the coupling-dependent scaling of $\eta$ in Eq.~\eqref{eq:hash_brown}, the $\kappa$-dependent correction to $g$ and $g_\times$ both scale as $O(g^0)$. This is in line with the notion that pinching-pole singularities come with terms of order $O(g^{-2})$, as expected from previous studies. Regardless of these contributions, finite-$\eta$ corrections appear in Eqs.~\eqref{eq:g_corr}--\eqref{eq:gc_corr} at order $O(g^4)$, thus leaving all of the above derivations unchanged. Another notable consequence of the pinching-pole singularities is that the coupling $g$ acquires an imaginary part. This suggests that the renormalised couplings are in general complex and that their RG flow should be studied in the space of complex, rather than real, numbers.

The quartic coupling are related by the following identities:
\begin{equation}
    \label{eq:nice_rel}
    g\equiv g^{(1)} = \bar{g}^{(2)}, \quad g_\times = 2 i \,\text{Im} g.
\end{equation}
The latter relation can also be derived independently both from  unitarity (see Section~\ref{sec:EFT_Unitarity} below), and from the absence of superficially pathological terms in two-point functions mentioned earlier. To see how, notice that from a perturbative diagrammatic point of view, any corrections to $g$---corresponding to some set of Feynman diagrams $\mathcal K$---can always be related to the $m^2$ corrections by joining two external lines that create an additional loop:
\begin{multline}
	\delta g[\mathcal K] \sim  
	\vcenter{\hbox{\begin{tikzpicture}
            \node[inner sep=0pt] (lu) at (-.4,.4) {};
            \node[inner sep=0pt] (ld) at (-.4,-.4) {};
            \node[inner sep=0pt] (lv) at (0,0) {};
            \draw
                (lu) edge[-,dotted] (lv)
                (ld) edge[-,dotted] (lv);
            \node[inner sep=0pt] (ru) at (2.2,.4) {};
            \node[inner sep=0pt] (rd) at (2.2,-.4) {};
            \node[inner sep=0pt] (rv) at (1.8,0) {};
            \draw
                (ru) edge[-,dotted] (rv)
                (rd) edge[-,dotted] (rv);
            \draw (.4,.4) arc [radius=.4, start angle=90, end angle= 270];
            \draw (1.4,.4) arc [radius=.4, start angle=90, end angle= -90];
            \draw[thick,fill=lightgray] (.4,-.5) rectangle (1.4,.5) node[pos=.5] {$\mathcal K$};
           \node[circle,draw=black,fill=black,inner sep=0pt,minimum size=4pt] (v) at (0,0) {};
           \node[circle,draw=black,fill=black,inner sep=0pt,minimum size=4pt] (v) at (1.8,0) {};
        \end{tikzpicture}}} \\ \Longrightarrow\quad \delta m^2[\mathcal K] \sim \vcenter{\hbox{\begin{tikzpicture}
            \node[inner sep=0pt] (lu) at (-.4,.4) {};
            \node[inner sep=0pt] (ld) at (-.4,-.4) {};
            \node[inner sep=0pt] (lv) at (0,0) {};
            \draw
                (lu) edge[-,dotted] (lv)
                (ld) edge[-,dotted] (lv);
            \draw (.4,.4) arc [radius=.4, start angle=90, end angle= 270];
            \draw (1.4,.4) arc [radius=.4, start angle=90, end angle= -90];
            \draw[thick,fill=lightgray] (.4,-.5) rectangle (1.4,.5) node[pos=.5] {$\mathcal K$};
           \node[circle,draw=black,fill=black,inner sep=0pt,minimum size=4pt] (v) at (0,0) {};
           \draw (1.8,0) arc [radius=.25, start angle=-180, end angle= 180];
           \node[circle,draw=black,fill=black,inner sep=0pt,minimum size=4pt] (v) at (1.8,0) {};
        \end{tikzpicture}}}\, .
\end{multline}
It follows that a corresponding set of diagrams $\mathcal K'$ correcting $g_\times$ must also exist:
\begin{multline}
	\delta g_\times[\mathcal K'] \sim  
	\vcenter{\hbox{\begin{tikzpicture}
            \node[inner sep=0pt] (lu) at (-.4,.4) {};
            \node[inner sep=0pt] (ld) at (-.4,-.4) {};
            \node[inner sep=0pt] (lv) at (0,0) {};
            \draw
                (lu) edge[-,dotted] (lv)
                (ld) edge[-,dotted] (lv);
            \node[inner sep=0pt] (ru) at (2.2,.4) {};
            \node[inner sep=0pt] (rd) at (2.2,-.4) {};
            \node[inner sep=0pt] (rv) at (1.8,0) {};
            \draw
                (ru) edge[-,dotted] (rv)
                (rd) edge[-,dotted] (rv);
            \draw (.4,.4) arc [radius=.4, start angle=90, end angle= 270];
            \draw[dashed] (1.4,.4) arc [radius=.4, start angle=90, end angle= -90];
            \draw[thick,fill=lightgray] (.4,-.5) rectangle (1.4,.5) node[pos=.5] {$\mathcal K'$};
           \node[circle,draw=black,fill=black,inner sep=0pt,minimum size=4pt] (v) at (0,0) {};
           \node[circle,draw=black,fill=white,inner sep=0pt,minimum size=4pt] (v) at (1.8,0) {};
        \end{tikzpicture}}} \\ \Longrightarrow\quad \delta m^2[\mathcal K'] \sim \vcenter{\hbox{\begin{tikzpicture}
            \node[inner sep=0pt] (lu) at (-.4,.4) {};
            \node[inner sep=0pt] (ld) at (-.4,-.4) {};
            \node[inner sep=0pt] (lv) at (0,0) {};
            \draw
                (lu) edge[-,dotted] (lv)
                (ld) edge[-,dotted] (lv);
            \draw (.4,.4) arc [radius=.4, start angle=90, end angle= 270];
            \draw[dashed] (1.4,.4) arc [radius=.4, start angle=90, end angle= -90];
            \draw[thick,fill=lightgray] (.4,-.5) rectangle (1.4,.5) node[pos=.5] {$\mathcal K'$};
           \node[circle,draw=black,fill=black,inner sep=0pt,minimum size=4pt] (v) at (0,0) {};
           \draw[dashed] (1.8,0) arc [radius=.25, start angle=-180, end angle= 180];
           \node[circle,draw=black,fill=white,inner sep=0pt,minimum size=4pt] (v) at (1.8,0) {};
        \end{tikzpicture}}} \, ,
\end{multline}
such that 
\begin{equation}
    \delta m^2[\mathcal K] + \delta m^2[\mathcal K'] = \text{finite}.
\end{equation}
Using Eq.~\eqref{eq:insp_eq} then establishes that 
\begin{equation}
    \delta g[\mathcal K] - \delta g_\times[\mathcal K'] = \text{finite}.
\end{equation}
In case the second term on the left-hand side consists of only divergent terms, as it actually does in the case of our one-loop calculation, this relation fixes $g_\times$ in terms of $g$, with the factor of 2, which already appears in Eqs.~\eqref{eq:corr1} and \eqref{eq:corr2}, coming from simple combinatorics.

\subsection{Wilsonian EFT at one loop and unitarity constraints}\label{sec:EFT_Unitarity}

To write down the one-loop Wilsonian effective action, we perform the usual rescaling of fields and momenta so that the partition function is expressed with the initial cutoff $\Lambda$. To this perturbative order, no wavefunction renormalisation is needed and therefore the scaling dimensions of operators have no anomalous contributions. Writing  $\Lambda' = \Lambda \zeta^{-1}$, with $\zeta>1$, the rescaled couplings are then given by
\begin{equation}
	\label{eq:running_couplings}
	\begin{gathered}
	m^2(\zeta) = \zeta^{2} (m^2 + \delta m^2(\zeta)), \\
	g(\zeta) = \zeta^{4-d} (g + \delta g(\zeta)), \\
    \bar g(\zeta) = \zeta^{4-d} (g + \delta \bar g(\zeta)), \\
    g_\times(\zeta) = \zeta^{4-d} \delta g_\times(\zeta).
	\end{gathered}
\end{equation}
The functions $\delta m^2(\zeta)$, $\delta g(\zeta)$, $\delta \bar g(\zeta)$ and $\delta g_\times(\zeta)$ are given, respectively, by Eqs.~\eqref{eq:m_corr} and \eqref{eq:g_corr}--\eqref{eq:gc_corr}. The full dissipative effective action was the one given by the expression in Eq.~\eqref{EFT_fin}, which we repeat here:
\begin{align}
       &S_{\rm eff}[\phi_1,\phi_2] = \int d^dx \, \bigg[ \frac{1}{2} \partial_\mu \phi_1 \partial^\mu \phi_1 - \frac{1}{2}m^2(\zeta) \phi_1^2 \nonumber\\
       & - \frac{1}{2} \partial_\mu \phi_2 \partial^\mu \phi_2 + \frac{1}{2}m^2(\zeta) \phi_2^2 - g(\zeta) \phi^4_1 + \bar g(\zeta) \phi^4_2 \nonumber\\
       &+ g_\times(\zeta) \phi^2_1 \phi_2^2 \bigg]. \nonumber
\end{align}
Given the presence of the pinching-pole singularities, one may wonder how meaningful such a one-loop effective action is. This is because, formally, perturbation theory breaks down and infinitely many higher-loop diagrams contribute at leading order. However, it turns out that, provided that the momentum shell we integrate out is thin enough, the usual perturbative picture is restored. 

To understand why this is the case, notice that perturbation theory in the usual sense breaks down as we perform the frequency integrals. This is because pairs of internal lines carrying the same momentum come with a factor of $g^{-2}$, which makes the notion of the loop order of a diagram different from its perturbative order. Concretely, this means that given the following two diagrams,
\begin{gather}
	 \vcenter{\hbox{\begin{tikzpicture}
            \node[inner sep=0pt] (lu) at (-.5,.5) {};
            \node[inner sep=0pt] (ld) at (-.5,-.5) {};
            \node[inner sep=0pt] (lv) at (0,0) {};
            \draw
                (lu) edge[-,dotted] (lv)
                (ld) edge[-,dotted] (lv);
            \node[inner sep=0pt] (ru) at (1.5,.5) {};
            \node[inner sep=0pt] (rd) at (1.5,-.5) {};
            \node[inner sep=0pt] (rv) at (1,0) {};
            \draw
                (ru) edge[-,dotted] (rv)
                (rd) edge[-,dotted] (rv);
            \draw (.5,-.5) arc [radius=.5, start angle=-90, end angle= 270];
           \node[circle,draw=black,fill=black,inner sep=0pt,minimum size=4pt] at (0,0) {};
           \node[circle,draw=black,fill=black,inner sep=0pt,minimum size=4pt] at (1,0) {};
        \end{tikzpicture}}} \sim \int_{\Lambda\zeta^{-1}}^{\Lambda} dp \, p^{d-2}I^{(1)}_\omega, \\[6pt]
        \vcenter{\hbox{\begin{tikzpicture}
            \node[inner sep=0pt] (lu) at (-.5,.5) {};
            \node[inner sep=0pt] (ld) at (-.5,-.5) {};
            \node[inner sep=0pt] (lv) at (0,0) {};
            \draw
                (lu) edge[-,dotted] (lv)
                (ld) edge[-,dotted] (lv);
            \node[inner sep=0pt] (ru) at (1.5,.5) {};
            \node[inner sep=0pt] (rd) at (1.5,-.5) {};
            \node[inner sep=0pt] (rv) at (1,0) {};
            \draw
                (ru) edge[-,dotted] (rv)
                (rd) edge[-,dotted] (rv);
            \draw (.5,-.5) arc [radius=.5, start angle=-90, end angle= 270];
           \node[circle,draw=black,fill=black,inner sep=0pt,minimum size=4pt] at (0,0) {};
           \node[circle,draw=black,fill=black,inner sep=0pt,minimum size=4pt] (k1) at (.5,.5) {};
           \node[circle,draw=black,fill=black,inner sep=0pt,minimum size=4pt] at (1,0) {};
           \node[circle,draw=black,fill=black,inner sep=0pt,minimum size=4pt] (k2) at (.5,-.5) {};
            \draw
                (k1) edge[-,bend right=30] (k2)
                (k2) edge[-,bend right=30] (k1);
        \end{tikzpicture}}} \sim \int_{\Lambda\zeta^{-1}}^{\Lambda} {\textstyle \prod_{i=1}^3} dp_i \, p_i^{d-2}I^{(3)}_\omega,
\end{gather}
different $I_\omega^{(i)}$ have the same perturbative order. If we, however, integrate only a thin shell of spatial momenta, integrating over them again suppresses higher loop diagrams. To be precise, introducing a logarithmic scale $s = \ln \zeta$, a diagram with $n$ loops will be suppressed by a factor of $s^n$. This suggests that we consider the coupling corrections \eqref{eq:m_corr} and \eqref{eq:g_corr}--\eqref{eq:gc_corr} as consistent and sensible only in the thin-shell limit. Nevertheless, the thin-shell limit is sufficient for deriving the $\beta$-functions, which we do below in Section~\ref{sec:beta}.

Before doing that, we discuss whether the EFT \eqref{EFT_fin} satisfies the necessary constraints imposed by the unitary time evolution of the microscopic dynamics. Indeed, any IR EFT derived from a unitary UV QFT must obey the following properties (see e.g.~Refs.~\cite{sasojanos1} and \cite{Crossley_eft,cgl_lectures}):
\begin{align}
    S_{\rm eff}[\phi_1,\phi_2] &= - \bar S_{\rm eff} [\phi_2,\phi_1] , \label{unitarity_1} \\
    \text{Im}\, S_{\rm eff}[\phi_1,\phi_2] & \geq 0 ,\label{unitarity_2}  \\
    S_{\rm eff}[\phi_1 = \phi_2] &= 0.\label{unitarity_3} 
\end{align}

As discussed above, while the running mass $m^2$ remains real to the perturbative order considered here (see Eq.~\eqref{eq:m_corr}), the running quartic coupling $g$ becomes complex and a new imaginary coupling $g_\times$ is generated. They are related by Eq.~\eqref{eq:nice_rel}. Note also that the quantum fields $\phi_{1,2}$ in the absence of any anomalous dimensions remain real. 

We now show explicitly that all three unitarity constraints \eqref{unitarity_1}--\eqref{unitarity_3} are satisfied by the EFT \eqref{EFT_fin}. First, by using relations between the quartic couplings \eqref{eq:nice_rel}, we find that the interacting part of the effective CTP Lagrangian density satisfies  
\begin{equation}
    \label{eq:int_part}
    \begin{gathered}
        \text{Re}\, \mathcal L_{\rm eff} \supset - \text{Re}g \left(\phi_1^4-\phi_2^4\right), \\
        \text{Im}\, \mathcal L_{\rm eff} \supset - \text{Im}g \left(\phi_1^2-\phi_2^2\right)^2.
    \end{gathered}
\end{equation}
From Eq.~\eqref{eq:g_corr} (note that $g\equiv g^{(1)})$, we see that $\text{Im}g<0$, which immediately ensures that the unitarity constraint \eqref{unitarity_2} is satisfied. Next, it is easy to see with the help of Eq.~\eqref{eq:int_part} that once we identify $\phi_1=\phi_2$, $S_{\rm eff} = 0$, thereby ensuring the constraint \eqref{unitarity_3}. At last, the constraint in Eq.~\eqref{unitarity_1} requires the real part of $S_{\rm eff}$ to be odd under the exchange $\phi_1 \leftrightarrow \phi_2$, while its imaginary part has to be even. It is straightforward to check that Eq.~\eqref{eq:int_part} implies both properties. We therefore conclude that our EFT \eqref{EFT_fin} with the coefficients given by Eqs.~\eqref{eq:m_corr} and \eqref{eq:g_corr}--\eqref{eq:gc_corr} (related by Eq.~\eqref{eq:nice_rel}) satisfies all necessary unitarity constraints. This provides an explicit example of the Schwinger-Keldysh EFT discussed in Ref.~\cite{sasojanos1}.

Finally, it is worth noting that another way to view the condition in Eq.~\eqref{unitarity_3} is as a constraint on the `classical' on-shell solutions of $\phi_{1,2}$. By investigating the two Euler-Lagrange equations of motion that follow from the effective action \eqref{EFT_fin}, i.e.,
\begin{align}
\partial^2 \phi_1 + m^2(\zeta) \phi_1 + 4 g(\zeta) \phi_1^3 - 2 g_\times(\zeta) \phi_2^2 \phi_1 &= 0 , \label{EL1} \\
\partial^2 \phi_2 + m^2(\zeta) \phi_2 + 4 \bar{g}(\zeta) \phi_2^3 + 2 g_\times(\zeta) \phi_1^2 \phi_2  &= 0 ,\label{EL2}
\end{align}
it is clear that because of the quartic coupling relation $g_\times = 2 i\, \text{Im} g$ (cf.~Eq.~\eqref{eq:nice_rel}), on shell, $\phi_1 = \phi_2 = \phi_{\rm on-shell}$, at which point $S_{\rm eff} = 0$. Despite the fact that our effective dissipative CTP action is complex, on shell, all imaginary parts of Eqs.~\eqref{EL1}--\eqref{EL2} cancel and $\phi_{\rm on-shell}$ remains real.   

\subsection{The $\beta$-functions of the EFT couplings}\label{sec:beta}

Finally, we now derive the explicit $\beta$-functions of the effective couplings $m^2$, $g$, $\bar g$ and $g_\times$ in the effective action \eqref{EFT_fin}. In terms of the running parameter $s = \ln \zeta$, the three running couplings are 
\begin{equation}
	\label{eq:log_running_couplings}
	\begin{gathered}
	m^2(s) = e^{2s} (m_0^2 + \delta m^2(s)), \\
	g(s) = e^{(4-d)s} (g_0 + \delta g(s)), \\
    \bar g(s) = e^{(4-d)s} (g_0 + \delta \bar g(s)), \\
    g_\times(s) = e^{(4-d)s} \delta g_\times(s),
	\end{gathered}
\end{equation}
where we have added the subscript $0$ to indicate the bare coupling of the original theory \eqref{eq:lagr_phi4}. The absence of a bare coupling $g_\times$ reflects the fact that the initial point of the RG flow is a closed, unitary theory.

To derive the $\beta$-functions, we use the `thin-shell' limit ($s\ll 1$) in which the all loop integrals simplify as
\begin{equation}
	\int_{\Lambda e^{-s}}^\Lambda dx \, f(x) = \Lambda f(\Lambda) s + O(s^2).
\end{equation}
Following this procedure and using Eqs.~\eqref{eq:m_corr} and \eqref{eq:g_corr}--\eqref{eq:gc_corr}, we obtain our final result:
\begin{align}
	\label{eq:bm}
\frac{\partial m^2}{\partial s} &= 2 m^2 + \frac{12 g \Omega_{d-2}}{(2\pi)^{d-1}} \frac{\Lambda^{d-1} \coth (\beta E_\Lambda/2)}{2 E_\Lambda}  + O(s), \\
	\label{eq:bg}
	\frac{\partial g}{\partial s} &= (4-d) g  -  \frac{36 g^2 \Omega_{d-2}}{(2\pi)^{d-1}} \frac{\Lambda^{d-1} \left[ \beta E_\Lambda + \sinh(\beta E_\Lambda) \right]}{8 E_\Lambda^3 \sinh^2(\beta E_\Lambda/2)}  \nonumber \\ 
    & - i \frac{36 \Omega_{d-2}}{(2\pi)^{d-1}} \frac{\Lambda^{d-1}}{8 E_\Lambda^2 \kappa \sinh^2(\beta E_\Lambda/2)} + O(s), \\
    \label{eq:bgb}
    \frac{\partial \bar g}{\partial s} &= (4-d) g  -  \frac{36 g^2 \Omega_{d-2}}{(2\pi)^{d-1}} \frac{\Lambda^{d-1} \left[ \beta E_\Lambda + \sinh(\beta E_\Lambda) \right]}{8 E_\Lambda^3 \sinh^2(\beta E_\Lambda/2)}  \nonumber \\ 
    & + i \frac{36 \Omega_{d-2}}{(2\pi)^{d-1}} \frac{\Lambda^{d-1}}{8 E_\Lambda^2 \kappa \sinh^2(\beta E_\Lambda/2)} + O(s), \\
    \label{eq:bgc}
	\frac{\partial g_\times}{\partial s} &= - i \frac{72 \Omega_{d-2}}{(2\pi)^{d-1}} \frac{\Lambda^{d-1}}{8 E_\Lambda^2 \kappa \sinh^2(\beta E_\Lambda/2)} + O(s).
\end{align}
The $\beta$-functions for each of these four couplings, which we collectively call $g_i = \{m^2, g, \bar g, g_\times\}$ is then given by $\beta_i \equiv \partial_s g_i$.\footnote{The $\beta$-functions are defined as logarithmic derivatives of the couplings with respect to the cutoff $\Lambda$. The relation between the cutoff $\Lambda$ and $s$ leads to $\Lambda \partial_\Lambda = \partial_s$, and, hence, to the definitions used in the Callan-Symanzik equation below.} The correction to the quartic couplings which arise as a result of our introduction of the thermal width $\kappa$ are the leading $O(g^0)$ terms in the respective $\beta$-functions. 

\section{Reduced complex space of couplings and criticality}\label{sec:beta_complex}

We now consider the existence of fixed points in the effective field theory \eqref{EFT_fin}. As noted above, to the perturbative order of our calculations, the anomalous dimension of $\phi$ vanishes. Hence, the Callan-Symanzik equation expressing the independence of a `physical' correlation function on the RG parameter $s$ reads
\begin{equation}
    \label{eq:callan_sym}
    \bigg[\frac{\partial}{\partial s} + \sum_{g_i} \beta(g_i) \frac{\partial}{\partial g_i} \bigg] G^{(n)}(s,g_i,p_1,...,p_n) = 0,
\end{equation}
where $G^{(n)}$ is a generic $n$-point correlation function in momentum space, which may include operators $\phi_1$ or $\phi_2$. The couplings $g_i$ runs over all four $m^2$, $g$, $\bar g$ and $g_\times$, and $\beta_i= \partial_s g_i$ follow from Eqs.~\eqref{eq:bm}--\eqref{eq:bgc}.

A fixed point (or a critical point) of the RG flow is defined as the point in the coupling space at which all $\beta$-functions vanish: $\partial_s m^2 = \partial_s g = \partial_s \bar g= \partial_s g_\times = 0$. At such points, correlators $G^{(n)}$ are independent of the RG scale $s$ when computed at the values of the renormalised couplings. A trivial fixed point can be found at $m^2,\, g\,,\bar g , g_\times = 0$. In this special instance, the UV theory \eqref{eq:lagr_phi4} is free and the spectral density \eqref{eq:free_spectral} does not receive any correction; thus, all loops vanish, in turn removing the third term from \eqref{eq:bg} and \eqref{eq:bgb}, and setting Eq.~\eqref{eq:bgc} to zero. Hence, at $g=0$, the only nontrivial equation is $\partial_s m^2|_{g=0} = 2 m^2$, from which we infer that $(m^2=0,\,g=0,\,\bar g=0,\,g_\times=0)$ is the standard Gaussian fixed point in any number of dimensions $d$. 

It is apparent that once we allow the UV theory to be interacting, i.e., to have the bare $g \neq 0$, then Eq.~\eqref{eq:bgc} can no longer be set to zero, which in the strict sense implies the absence of any interacting fixed points, at least for the current choice of initial conditions to the flow (that is, requiring the UV theory to be unitary). We comment on the possibility of having different initial conditions in the conclusion. Nevertheless, by restricting the analysis to correlation functions constructed only out of a single type of field, $\phi_1$ or $\phi_2$, we still find that certain such correlators do behave as if at a fixed point for special values of the coupling. This is because their diagrammatic expressions do not contain either the cross-coupling $g_\times$, or the complex conjugate of the appropriate quartic self-coupling. We denote such single-time-contour correlators by $\bar G_{s}^{(n)}$. They include the time-ordered Feynman correlators computed at `tree-level' of the EFT, which enter into the computation of scattering amplitudes and the S-matrix. Note, of course, that the `tree-level' EFT is a quantum action that includes various loop corrections computed from the original action \eqref{eq:lagr_phi4}. 

An independent way to view such critical points is from the perspective of an open QFT. In particular, we could imagine some fine-tuned external source that would drive the theory in manner that would decouple the two legs of the CTP contour. A fixed point in the reduced coupling space could then potentially correspond to a fixed point is such an open QFT.

Let us now focus on type-1 correlators for concreteness; for those, by definition, $\partial \bar G^{(n)} / \partial \bar g = \partial \bar G^{(n)} / \partial g_\times = 0$, so that the Callan-Symanzik equation \eqref{eq:callan_sym} will only depend on $m^2$ and $g$. This defines our `reduced space of couplings' in which it is then possible to consider the existence of interacting critical points for the single-time-contour correlators $\bar G^{(n)}$. Such critical point are defined by $\partial_s m^2 = \partial_s g = 0$ at $g\neq 0$, with $\partial_s \bar g \neq 0$ and $\partial_s g_\times \neq 0$.

For purposes of this analysis, it is convenient to make all involved quantities dimensionless. Explicitly, we define
\begin{equation}
    \begin{gathered}
        m^2 = \tilde m^2 \Lambda^2, \\
        g = \tilde g \Lambda^{4-d}, \\
        \beta = \tilde \beta \Lambda^{-1}, \\ 
        \kappa = \tilde \kappa E_\Lambda \Lambda^{2(d-4)},
    \end{gathered}
\end{equation}
where the quantities with the overhead tildes are dimensionless. Introducing $\alpha_d = \Omega_{d-2}/(2\pi)^{d-1}$, Eqs.~\eqref{eq:bm} and \eqref{eq:bg} then become
\begin{align}
    \label{eq:adim_bm}
\partial_s \tilde m^2 &= 2 \tilde m^2 + 6 \alpha_d \tilde g \frac{\coth{(\tilde \beta \sqrt{1 + \tilde m^2}/2})}{\sqrt{1 + \tilde m^2}}, \\ 
\partial_s \tilde g &= (4-d) \tilde g  \label{eq:adim_bg} \\ 
&- \frac{9\alpha_d}{2} \frac{\tilde g^2\spars{\tilde \beta \sqrt{1+\tilde m^2} + \sinh(\tilde \beta \sqrt{1+\tilde m^2}) } + i \tilde\kappa^{-1}}{(1+\tilde m^2)^{3/2} \sinh^2(\tilde \beta \sqrt{1+\tilde m^2}/2)} . \nonumber
\end{align}
The fixed points are solutions to $\partial_s \tilde m^2 = \partial_s \tilde g =  0$. We will first analyse them numerically and then also analytically at low temperature. 

\begin{figure}[t!]
    \centering
    \includegraphics[width=\linewidth]{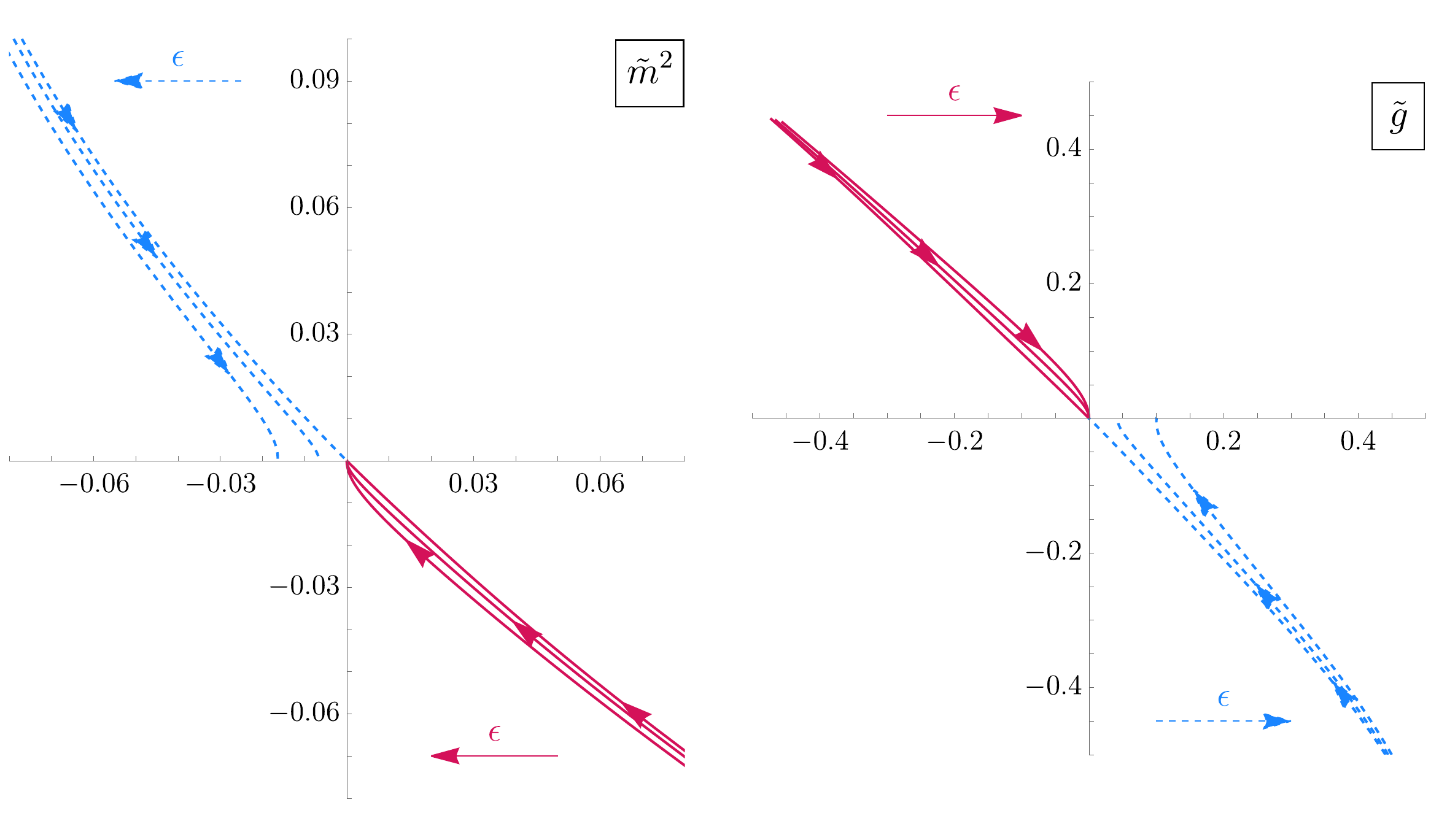}
    \caption{Reduced complex coupling space fixed points in $d=4-\epsilon$ at fixed $\tilde\kappa=1$ and increasing $\tilde\beta$, plotted for $\epsilon=0,\,0.02,\,0.05$. The solid magenta line corresponds to $(\tilde m^2_*,\tilde g_*)$. The blue dashed line corresponds to $(\tilde m^2_\circ,\tilde g_\circ)$. Arrows points in the direction of decreasing temperature and the curves are plotted for increasing value of $\epsilon$ (see the $\epsilon$-arrows).}
    \label{fig:fps_4etad}
\end{figure}

\textbf{Numerical results.---}We show the result of the numerical analysis in Figure~\ref{fig:fps_4etad}. At fixed values of the temperature and the thermal width (treated as an independent parameter), we find two interacting fixed points, which we denote by $(\tilde m^2_*, \tilde g_*)$ and $(\tilde m^2_\circ, \tilde g_\circ)$.

The behaviour at the critical dimension $d=4$ and slightly below is qualitatively different. At $d=4$, as the temperature is lowered (i.e., for increasing $\tilde\beta$), the two branches collapse to the origin of the $\mathbb C^2$ space of couplings $\tilde m^2$ and $\tilde g$---i.e., to the Gaussian fixed point---in an approximately symmetric manner. However, when $d<4$, only one of the two fixed points $(\tilde m^2_*,\tilde g_*)$ runs to the Gaussian fixed point at low temperatures, whereas $(\tilde m^2_\circ,\tilde g_\circ)$ tends to an $\epsilon$-dependent point in $\mathbb C^2$. The low-temperature picture is somewhat reminiscent of the flow diagram of the Euclidean case in $d=4-\epsilon$, where a flow connecting the Gaussian fixed point to the Wilson-Fisher fixed point exists. It is therefore natural to wonder whether in our finite temperature setup, this analogy with the Wilson-Fisher fixed point is appropriate even precisely at $d=4$. We will treat $(\tilde m^2_\circ,\tilde g_\circ)$ as a candidate for a novel Wilson-Fisher-type fixed point in the reduced coupling space.

Let us now study the linearised RG flow in the vicinity of the two fixed points. We write
\begin{equation}
    \begin{gathered}
         \tilde m^2 = \tilde m^2_{\smhash} + \delta \tilde m^2, \\
        \tilde g = \tilde g_{\smhash} + \delta \tilde g,
    \end{gathered}
\end{equation}
where $(\tilde m^2_{\smhash},\tilde g_{\smhash})$ can represent either one of the two fixed points $(\tilde m^2_*, \tilde g_*)$ or $(\tilde m^2_\circ, \tilde g_\circ)$. Then,
\begin{equation}
    \label{eq:lin_flow}
	\partial_s \begin{pmatrix} \delta \tilde m^2 \\ \delta \tilde g \end{pmatrix} = \begin{pmatrix} a_{11} & a_{12} \\ a_{21} & a_{22} \end{pmatrix} \begin{pmatrix} \delta \tilde m^2 \\ \delta \tilde g \end{pmatrix},
\end{equation}
where the $a_{ij}$ coefficients are the order $O(1)$ prefactors appearing in the linear term of the expanded Eqs.~\eqref{eq:adim_bm} and \eqref{eq:adim_bg}. Their are rather lengthy so We do not state them here explicitly. Importantly, the coefficients do depend on the fixed point at which they are evaluated. The eigenvalues of $a_{ij}$, which we name $\Delta_{m_{\smhash}}$ and $\Delta_{g_{\smhash}}$, encode the scaling dimensions of deformations away from the corresponding fixed point; each eigenvalue with a positive real part corresponds to a relevant deformation. 

\begin{figure}[t!]
    \centering
    \includegraphics[width=\linewidth]{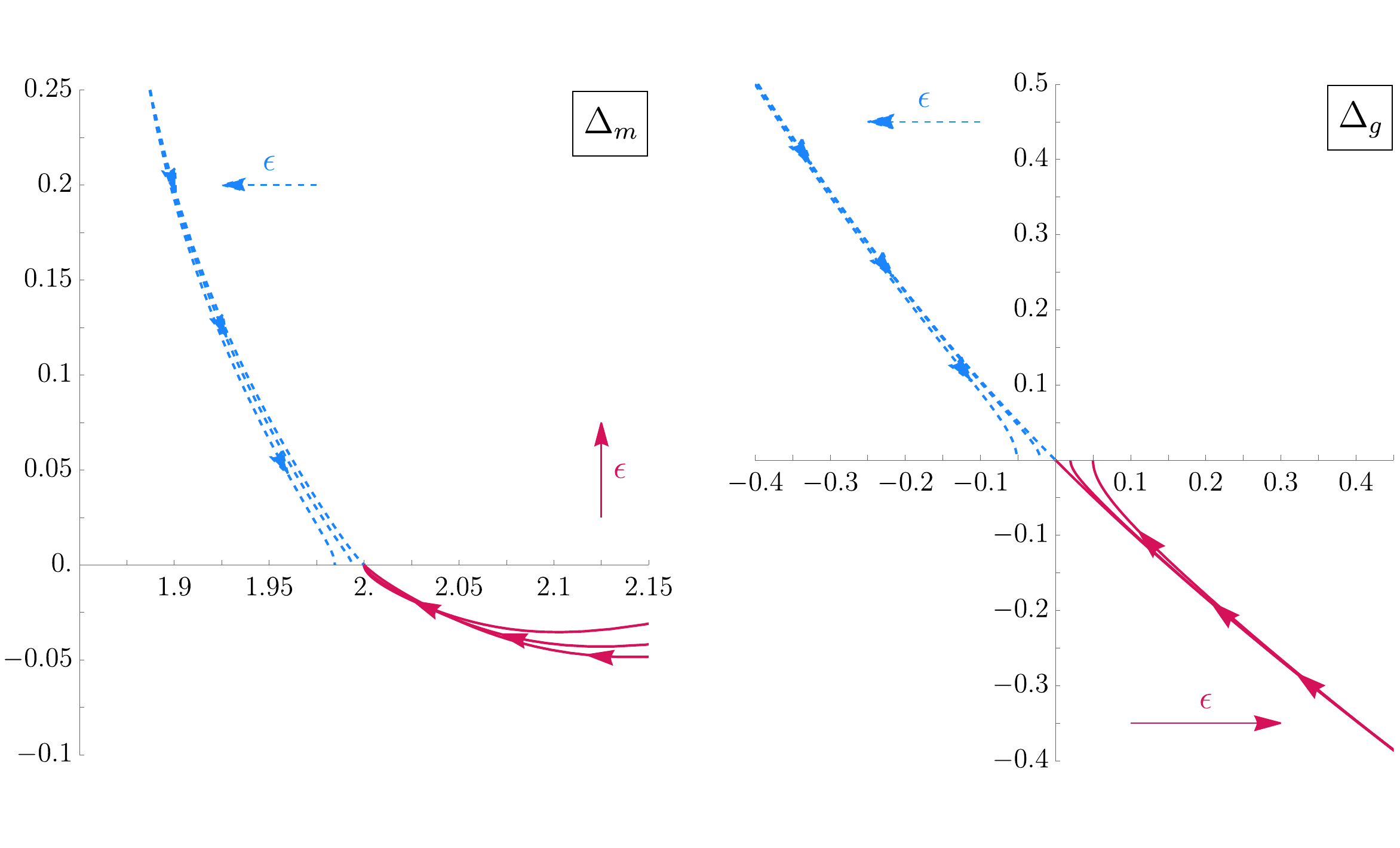}
    \caption{Eigenvalues of the linearised flow around the $d=4-\epsilon$ fixed points shown in Figure \ref{fig:fps_4etad}.}
    \label{fig:eigs_4etad}
\end{figure}

We plot numerical results for the scaling dimensions of deformations at the fixed points in Figure~\ref{fig:eigs_4etad}. We find out that at all temperatures and for numbers of dimensions $d=4-\epsilon$ we have been discussing, $(\tilde m^2_\circ, \tilde g_\circ)$ has one relevant and one irrelevant deformation,
\begin{equation}
    \text{Re}\pars{\Delta_{m_\circ}}>0, \quad \text{Re}\pars{\Delta_{g_\circ}}<0.
\end{equation}
It is important to stress that this remains true even at the critical dimension, $d=4$. On the other hand, we find that
\begin{equation}
    \text{Re}\pars{\Delta_{m_*}}>0, \quad \text{Re}\pars{\Delta_{g_*}}>0.
\end{equation}

\textbf{Analytic approximation.---}In order to better understand the nature of these fixed points, we continue with an analytic treatment. By using the numerical results, notice that in all cases considered, $\tilde m^2_{\smhash}\ll 1$ in the $\tilde\beta\gg1$ regime. At low temperatures, 
\begin{equation}
    \coth\pars{\frac{\tilde \beta \sqrt{1+\tilde m^2}}{2}} = 1 + O\pars{e^{-\tilde\beta\sqrt{1 + \tilde m^2}}},
\end{equation}
which can be used to simplify Eq.~\eqref{eq:adim_bm}, giving
\begin{equation}
    \partial_s \tilde m^2 \approx 2 \tilde m^2 + \frac{6 \alpha_d \tilde g}{\sqrt{1 + \tilde m^2}}.
\end{equation}
This signals that a fixed point must exhibit the following relative scaling: $\tilde m^2 \sim \tilde g$. One can then expand in $\tilde m^2$ to find corrections up to the perturbative order of the RG flow equations. Using $d=4-\epsilon$, we find the approximate low-temperature RG equations to be
\begin{align}
\partial_s \tilde m^2 &= 2 \tilde m^2 + 6 \alpha_{4-\epsilon} \tilde g, \\
\partial_s \tilde g &= \epsilon \tilde g \\ 
&- \frac{9 \alpha_{4-\eta}}{2} \bigg[2 \tilde g^2 + \frac{4 i e^{-\tilde \beta}}{\tilde\kappa} \bigg( 1 - \frac{3}{2}\tilde m^2 + \frac{15}{8} \tilde m^4\bigg)\bigg]. \nonumber
\end{align}

Let us first consider the resulting fixed point in the critical number of spacetime dimensions, $d=4$, with $\epsilon = 0$. To leading order in temperature, these two fixed points exist at complex values of the couplings: 
\begin{equation}
    \label{eq:fp_approx}
    \begin{gathered}
        \tilde m^2_* = \frac{3(1-i)e^{-\tilde\beta/2}}{4\pi^2\tilde\kappa^{1/2}}, \quad \tilde g_* = \frac{(i-1)e^{-\tilde\beta/2}}{\tilde\kappa^{1/2}}, \\
        \tilde m^2_\circ = - \tilde m^2_*, \quad \tilde g_\circ = - \tilde g_*.
    \end{gathered}
\end{equation}
Interestingly, the dependence of the novel fixed points on $\tilde\beta$ is non-analytic, as $\tilde m^2_*, \, \tilde g_* \sim e^{-\Lambda/T}$. 

Next, we study the two fixed points at finite $\epsilon$. By investigating the numerical results, it is clear that one cannot built a consistent perturbative expansion in $\epsilon$ around the results in Eq.~\eqref{eq:fp_approx}. The reason for this is that the limits of $T\to 0 $ and $\epsilon \to 0$ do not commute, so the appropriate scaling we require to find the fixed points is $e^{-\tilde\beta/2} \ll \epsilon \ll 1$. Indeed, Figure~\ref{fig:eigs_4etad} shows that the endpoints of our numerical curves are at $\tilde\beta = O(10)$, for which $e^{-\tilde\beta/2}= O(10^{-5})$, which occurs for example around $\epsilon = O(10^{-2})$. Taking the ansatz
\begin{equation}
    \label{eq:ceta_exp2}
    \tilde m^2 \approx \epsilon \tilde m_{(1)}^2, \quad \tilde g \approx \epsilon \tilde g_{(1)}, 
\end{equation}
we then find that to leading order in temperature and $\epsilon$,
\begin{equation}
    \begin{gathered}
        \tilde m^2_* = 0, \quad \tilde g_* = 0, \\
        \tilde m^2_\circ = -\frac{\epsilon}{3}, \quad \tilde g_\circ = \frac{2 \pi^2}{9} \epsilon, 
    \end{gathered}
\end{equation}
which agrees with our numerical findings and the literature.\footnote{The Wilson-Fisher fixed point calculated in Euclidean field theory actually has $m_\circ^2 = - \epsilon/6$. This `discrepancy' can be traced to a different prescription for the definition of the cutoff. In particular, in the Euclidean calculation, one imposes the cutoff on frequencies and momenta, which in terms of our analysis corresponds to changing $\alpha_d\to\alpha_{d+1}$. Importantly, this difference does not affect the critical exponents, which are the important physical quantities.} As a further consistency check, we can also calculate approximate eigenvalues of the linearised problem. Diagonalising the matrix from Eq.~\eqref{eq:lin_flow}, and expanding the results in $\epsilon$ and then in $T$, we find that 
\begin{align}
       \!\!\! \Delta_{m_*} \! &= 2 + O(e^{-\tilde\beta/2}), \!\! & \Delta_{g_*} \! &= \epsilon + O(e^{-\tilde\beta/2}), \\
       \!\!\! \Delta_{m_\circ} \! &= 2-\frac{\epsilon}{3} + O(e^{-\tilde\beta/2}),\!\! & \Delta_{g_\circ} \!&= - \epsilon + O(e^{-\tilde\beta/2}),
\end{align}
which are the expected scaling dimensions of deformations of, respectively, the Gaussian fixed point and the Wilson-Fisher fixed point. They agree with the numerical results from Figure~\ref{fig:eigs_4etad}. 

\textbf{Summary.---}We have demonstrated the existence of two novel (dynamical) interacting fixed points, $(\tilde m^2_*, \tilde g_*)$ and $(\tilde m^2_\circ, \tilde g_\circ)$, in the reduced coupling space of the EFT on the CTP contour. While these are not fixed points in the usual sense and in the full EFT \eqref{EFT_fin} due to the seemingly inevitable presence of the cross-interaction that couples $\phi_1$ and $\phi_2$, these novel fixed points can nevertheless be `experienced' by the time-ordered correlation functions computed at the `tree-level' order of the EFT defined on a single leg of the CTP contour. Such correlators are typically associated with transition amplitudes between the in- and out-vacuum states. Hence, the fixed points found here therefore provide a generalised notion of the standard Gaussian and Wilson-Fisher fixed points. Importantly, they exist in the integer (critical) number of spacetime dimension $d=4$. 

\begin{figure}[t!]
    \centering
    \includegraphics[width=\linewidth]{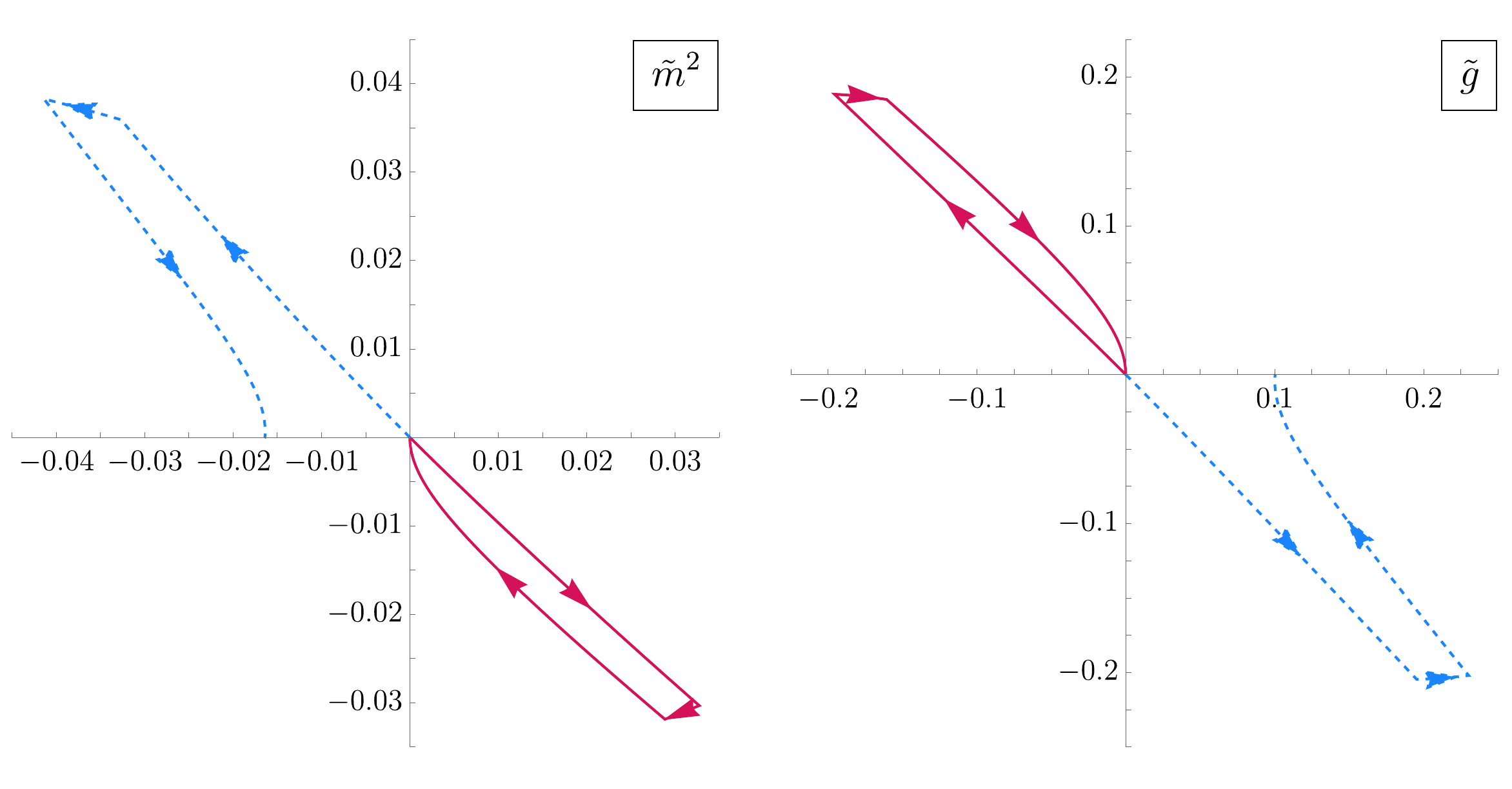}
    \caption{A diagram depicting the behaviour of the two fixed points. The three segments of each curve following the arrows are: increasing $T$ at $d=4$, fixed $T$ and decreasing $d = 4-\epsilon$, fixed $\epsilon > 0$ and decreasing $T\to 0$. The solid magenta curve ends at the Gaussian fixed point when $\epsilon >0$ and $T \to 0$. The dashed blue curve ends at the standard Wilson-Fisher fixed point when $\epsilon >0$ and $T \to 0$.}
    \label{fig:heat_engine}
\end{figure}

For a way to think about our results and how they fit into the usual picture of fixed points in $\phi^4$ theory, consider the $T=0$ theory in $d=4$, which only has the Gaussian fixed point. By putting the theory in a thermal state, the Gaussian fixed point splits into two new fixed points, with their temperature dependence being non-analytic (as per Eq.~\eqref{eq:fp_approx}). At some finite $T$, one can then imagine keeping the temperature fixed while continuously decreasing the number of dimensions $d = 4 -\epsilon$, tuning $\epsilon$ away from $0$. At some finite fixed $\epsilon>0$, the theory can again be `cooled' to the vacuum state with $T = 0$. This process, which clearly demonstrates the non-commuting nature of the $T\to 0$ and $\epsilon \to 0$ limits, draws our two continuous curves in the complex space of reduced couplings, one ending at the Gaussian fixed point, the other at the Wilson-Fisher fixed point. Note that at $T = 0$, $g = \bar g$ and $g_\times = 0$, so the endpoint is a genuine fixed point of the full theory for all correlators. We show this non-commuting flow of the fixed points in Figure~\ref{fig:heat_engine}. It is therefore natural to argue that the $d=4$ branch of the blue dashed curve in Figure should be regarded as a generalised Wilson-Fisher fixed point, which exists for time-ordered (or single CTP branch) correlators at exactly $d=4$ ($\epsilon=0$) in the thermal CTP effective theory \eqref{EFT_fin} derived from the $\phi^4$ theory in Eq.~\eqref{eq:lagr_phi4}.

A natural outstanding question is whether one can associate the complex thermal scaling dimensions $\Delta_m$ and $\Delta_g$ in $d=4$ (presented in Figure~\ref{fig:eigs_4etad}) with experimentally measurable critical exponents, normally computed from statistical field theory. While we do not resolve this question, we do offer a few comments here. Let us assume that it is possible (and well-defined) to analytically continue our EFT time-ordered correlators composed of $\phi_1$ (for which the dynamical `reduced-space' critical points exist) to Euclidean correlators (see for example \cite{DolanJackiw}). Then there naively exists a direct connection between the complex $\Delta_m$ and $\Delta_g$ and the usual, now also complex, critical exponents. For example, $\nu = 1/\Delta_m$. Since the underlying theory is dissipative, this finding seems then compatible with the notion that such fixed points would be non-unitary. Nevertheless, a significantly more thorough examination of such (open) Euclidean fixed points and relations between critical exponents will be required to make any statement precise. Regardless, however, we end this section by noting that the main purpose of the CTP formalism is to gain access to time-dependent observables, so a discussion of an analytic continuation to Euclidean time seems somewhat counterproductive and simpler methods may be available for the study of complex (open or dissipative) statistical critical exponents.  

\section{Conclusion}
\label{sec:conc}

In this work, we explicitly derived a dissipative EFT of the type proposed in Ref.~\cite{sasojanos1}. The EFT exhibits an RG-induced `influence functional' with a local coupling between the two legs of the CTP time contour. Its appearance can be traced to the presence of pinching-pole singularities, which typically appear in QFTs at finite temperature, and to the procedure we used to regularise such singularities, which involves introducing a thermal width for the elementary particles, as expected at $T\neq0$. We then proceeded to analyse the critical properties of a `reduced' RG flow exhibited by the correlators defined on a single leg of the time contour, computed to perturbative order in the couplings which allow for the EFT to be treated classically (at tree-level). We uncovered two novel interacting fixed points, which could be understood as thermal CTP deformations of the Gaussian and the Wilson-Fisher fixed points in integer number of spacetime dimensions, $d=4$. 

There are a number of obvious and important extensions to our work. One is to better understand higher-order perturbative corrections and to clarify precisely all possible effective couplings that may appear in the IR EFT. It would also be beneficial to the overall picture presented here to calculate the thermal width (and $\kappa$) from the underlying QFT and not treat it as a phenomenological, regularisation-motivated parameter. Moreover, it would be illuminating to better understand the connection between our procedure and non-perturbative treatments, such as the exact or functional RG techniques (for a review of such methods, see Ref.~\cite{Polonyi:2001se}).

It would also be interesting to understand the connection between our results and the behavior of putative open and/or non-unitary field theories. One could for instance wonder whether allowing the bare theory to be non-unitary, for instance by including cross-branch couplings, would lead to the discovery of critical points in the full space of complex couplings. Non-unitary theories evade the restrictions imposed by the $c$-theorem \cite{zamolodchikov:Ctheo} and its higher-dimensional generalisations, thereby opening the door to unconventional RG flows, such as the ones recently discussed in Refs.~\cite{JKP_rg1,JP_rg2}. Beyond a mathematical analysis of such fixed points, the relevance of their existence for real-life systems also remains an open question. For example, it has long been known (see e.g.~Ref.~\cite{Nauenberg_1975}) that certain models can exhibit critical properties with complex scaling dimensions. However, since they only possess discrete, rather than full, scale invariance, they do not feature in the description of stable critical phases. This type of a fixed point can be successfully employed to describe unstable systems, which can undergo catastrophic first order phase transitions, such as earthquakes \cite{sornette:jpa-00247086}. Finally, further future development of the CTP techniques to study novel (actual or `reduced-space') fixed points should also be of relevance to condensed matter systems, thermal high-energy states and strongly coupled states matter (e.g., in open systems, see Refs.~\cite{Jana:2020vyx,Loganayagam:2022zmq}) that can be described by using holographic techniques. 

\begin{acknowledgements}
We would like to thank Doğan Akpinar, Carl Bender, Felix Haehl, Janos Polonyi, Justin Kulp, Neil Turok, Mile Vrbica and Matthew Walters for illuminating or otherwise inspiring conversation had during the completion of this work. The work of G.F. is supported by an Edinburgh Doctoral College Scholarship (EDCS). The work of S.G. was supported by the STFC Ernest Rutherford Fellowship ST/T00388X/1. The work is also supported by the research programme P1-0402 and the project N1-0245 of Slovenian Research Agency (ARIS).
\end{acknowledgements}

\appendix
\section{Regularised delta function identities}\label{app:delta}

The analysis in this work uses a few relations involving the composition and other properties of regularised distributions. We state the relevant ones here.

We start by deriving Eq.~\eqref{eq:niemi}, which is a straightforward application of Eq.~\eqref{eq:reg_delta}. In particular,
\begin{multline}
    \frac{\deps(x^2-\alpha^2)}{x^2-\alpha^2 \pm i \eta} = \frac{1}{\pi} \frac{\eta \spars{(x^2-\alpha^2) \mp i \eta}}{\spars{(x^2-\alpha^2)^2 + \eta^2}^2} = \\ = \frac{1}{\pi} \frac{\eta (x^2-\alpha^2)}{\spars{(x^2-\alpha^2)^2 + \eta^2}^2} \mp i \pi \spars{\frac{1}{\pi} \frac{\eta}{(x^2-\alpha^2)^2 + \eta^2} }^2.
\end{multline}
The second term is the square of $\deps$ and the first term can be rewritten as its derivative by observing that
\begin{equation}
    \partial_{x^2} \deps(x^2 - \alpha^2) = -\frac{1}{\pi} \frac{2 (x^2 - \alpha^2)}{\spars{(x^2-\alpha^2)^2 + \eta^2}^2}. \label{der_delta_a2}
\end{equation}
Inserting \eqref{der_delta_a2} into the expression above gives Eq.~\eqref{eq:niemi}. 

Another important relation is in Eq.~\eqref{eq:hash_brown}, which is crucial in regularising the diverging integral \eqref{eq:ill_part}. In order to make such an identification, one needs to prove that a certain regularised version of the free spectral density \eqref{eq:free_spectral} could be written in form of the weakly interacting result \eqref{eq:int_spectral}. To achieve the goal one could try to apply the standard composition rule for $\delta$ functions,
\begin{equation}
	\label{eq:equal?}
	\deps(x^2-\alpha^2) \stackrel{?}{=} \frac{1}{2\alpha} \pars{ \deps(x-\alpha) + \deps(x+\alpha)}.
\end{equation}
At finite $\eta$, this expression is not as straightforward as it seems. In particular, $\eta$ has the same dimension as the argument of $\deps$, hence, the infinitesimal quantities on the two sides cannot be the same. Albeit, they must be proportional to each other. To understand this, we write
\begin{equation}
	\deps(x\pm \alpha) = \frac{1}{\pi} \frac{\tilde\eta}{(x\pm\alpha)^2 + \tilde\eta^2}.
\end{equation}
To uncover the relationship between $\eta$ and $\tilde\eta$, one can then calculate an integral of the form in \eqref{eq:test_integral} using both expressions for the $\deps$, impose the leading terms to be identical and check how much the subleading terms differ. One finds that the identification
\begin{equation}
	\eta = 2 \alpha \tilde\eta,
\end{equation}
which we used in the main body of the paper, makes the integrals calculated with two different prescriptions equal up to order $O(\eta^2)$, which in terms of the coupling $g$ corresponds to $O(g^4)$. Thus Eq.~\eqref{eq:hash_brown} can be used to the perturbative order considered in this work.

\section{Higher pinching-pole singularities and regularisation}
\label{app:higher_PPs}

In this appendix, we describe in more detail the mathematical groundwork beneath the regularisation adopted in the main text. In particular, we derive a formula whose special case allows us to write Eq.~\eqref{eq:ill_part} as \eqref{eq:ill_part2}. 

Consider the integral
\begin{equation}
	\label{eq:test_integral}
	I_\text{test} = \int_{\mathbb R} dx \, f(x) [\deps(x^2-\alpha^2)]^n,
\end{equation}
where $n\geq1$. We assume $f$ to be a test function that is well-behaved at complex infinity, which allows us to compute \eqref{eq:test_integral} using a contour integral with a contour closed either in the upper or lower half plane. We choose the former. Moreover, we assume that the poles of $f$ do not overlap with the ones from $\deps$. Then, using Eq.~\eqref{eq:reg_delta}, we find that for finite $\eta$,
\begin{align}
	I_\text{test} &= 2 \pi i  \Big[ \text{Res}(f \deps^n, \sqrt{\alpha^2 + i \eta}) \nonumber \\  
    &+ \text{Res}(f\deps^n, -\sqrt{\alpha^2 - i \eta}) + \ldots \Big],
\end{align}
where the dots represent the contributions coming from the poles of $f$ in the upper half plane, which are parametrically suppressed by a factor $\eta^n$, so that they become irrelevant in the $\eta\to0$ limit that we are interested in. Each of the poles is of order $n$, hence, the residues can be computed with the usual formula:
\begin{equation}
	\text{Res}(g,z_0) = \frac{1}{(n-1)!} \lim_{z\to z_0} \partial_z^{n-1}[ (z-z_0)^n g(z)]. 
\end{equation}
Upon taking the limit of $\eta\to0$, this gives an approximate representation of each residue in terms of a power series in $\eta$. The leading term of each of the residues is
\begin{align}
	&\text{Res}(f\deps^n, \sqrt{\alpha^2 + i \eta}) = \nonumber\\
    &\quad= - i \frac{f(\alpha)}{\alpha (4\pi)^n} \frac{(2n-2)!}{[(n-1)!]^2} \frac{1}{\eta^{n-1}} (1 + O(\eta^2)),\\
	&\text{Res}(f\deps^n, -\sqrt{\alpha^2 - i \eta}) = \nonumber\\
    &\quad= - i \frac{f(-\alpha)}{\alpha (4\pi)^n} \frac{(2n-2)!}{[(n-1)!]^2} \frac{1}{\eta^{n-1}} (1 + O(\eta^2)).
\end{align}
The full integral is then given by 
\begin{equation}
	\label{eq:fint_eps}
	I_\text{test} = \frac{2 \pi (f(\alpha) + f(-\alpha))}{\alpha (4\pi)^n} \frac{(2n-2)!}{[(n-1)!]^2} \frac{1}{\eta^{n-1}} (1 + O(\eta^2)),
\end{equation}
where the $O(\eta^2)$ part can be determined order-by-order. This is the formula we used to derive Eq.~\eqref{eq:ill_part2}.

We conclude with a comment on the analytic structure of the integrand. In the $\eta\to0$ limit, complex conjugate poles pinch the real line; this is called a \textit{pinch singularity} \cite{Eden_anSmatrix}. Pinch singularities are not always a problem as, for example, the $n=1$ case gives a finite result. For divergences to arise, one needs to have $n\geq2$. Those integrals do naturally appear in the calculation of Feynman diagrams with pairs of equal loop momenta. As they can be generated by taking the limit of vanishing external momentum, which makes different poles in the complex frequency plane collide, these are called these are called \textit{pinching-pole singularities}. 

The pinching-pole singularities are therefore a subset of pinch singularities, which a standard construction associates with branch points (see Chapter 2 of Ref.~\cite{Eden_anSmatrix}). One could then wonder whether the divergence in \eqref{eq:fint_eps} could be tamed by going to a different Riemann sheet. This does not work as the discontinuity across the cut is
\begin{align}
	\text{Disc}(I_\text{test}) &= 2 \pi i  \Big[ \text{Res}(f \deps^n, \sqrt{\alpha^2 + i \eta}) \nonumber \\  
    &+ \text{Res}(f\deps^n, \sqrt{\alpha^2 - i \eta}) \Big],
\end{align}
which, if used to remove the singularity, makes the integral vanish.

For comparison, consider a case analogous to when two pinch singularities are kept separate by not setting the external momentum to zero. In particular, this would give the integral of the form 
\begin{equation}
	I_\text{test}' = \int_{\mathbb R} dx \, f(x) \deps(x^2-\alpha^2) \deps((x-x_*)^2-\alpha^2).
\end{equation}
The $p\to0$ limit in this case corresponds to taking $x_*\to0$. Rather than having four second-order poles at $\pm\sqrt{\alpha^2 \pm i \eta}$, this expression has eight simple poles: four in the same location and four at $x_* \pm\sqrt{\alpha^2 \pm i \eta}$. Closing the contour in the upper half plane, one finds that
\begin{align}
	\label{eq:second_test_fin}
	I_\text{test}' &= \frac{\eta}{2 \pi \alpha x_*^2} \left[\frac{f(\alpha)+f(x_* - \alpha)}{(x_* -2 \alpha)^2} \right.\nonumber\\
    &\left.+\frac{f(-\alpha) + f(x_* + \alpha)}{(x_* +2 \alpha)^2} \right] + O(\eta^2).
\end{align}
This shows that at finite external momentum, diagrams which give rise to pinching-pole singularities in the $p\to0$ limit are suppressed by higher powers of $\eta$ and can therefore be neglected.

\bibliography{refs}

\begin{thebibliography}{55}%
\makeatletter
\providecommand \@ifxundefined [1]{%
 \@ifx{#1\undefined}
}%
\providecommand \@ifnum [1]{%
 \ifnum #1\expandafter \@firstoftwo
 \else \expandafter \@secondoftwo
 \fi
}%
\providecommand \@ifx [1]{%
 \ifx #1\expandafter \@firstoftwo
 \else \expandafter \@secondoftwo
 \fi
}%
\providecommand \natexlab [1]{#1}%
\providecommand \enquote  [1]{``#1''}%
\providecommand \bibnamefont  [1]{#1}%
\providecommand \bibfnamefont [1]{#1}%
\providecommand \citenamefont [1]{#1}%
\providecommand \href@noop [0]{\@secondoftwo}%
\providecommand \href [0]{\begingroup \@sanitize@url \@href}%
\providecommand \@href[1]{\@@startlink{#1}\@@href}%
\providecommand \@@href[1]{\endgroup#1\@@endlink}%
\providecommand \@sanitize@url [0]{\catcode `\\12\catcode `\$12\catcode `\&12\catcode `\#12\catcode `\^12\catcode `\_12\catcode `\%12\relax}%
\providecommand \@@startlink[1]{}%
\providecommand \@@endlink[0]{}%
\providecommand \url  [0]{\begingroup\@sanitize@url \@url }%
\providecommand \@url [1]{\endgroup\@href {#1}{\urlprefix }}%
\providecommand \urlprefix  [0]{URL }%
\providecommand \Eprint [0]{\href }%
\providecommand \doibase [0]{https://doi.org/}%
\providecommand \selectlanguage [0]{\@gobble}%
\providecommand \bibinfo  [0]{\@secondoftwo}%
\providecommand \bibfield  [0]{\@secondoftwo}%
\providecommand \translation [1]{[#1]}%
\providecommand \BibitemOpen [0]{}%
\providecommand \bibitemStop [0]{}%
\providecommand \bibitemNoStop [0]{.\EOS\space}%
\providecommand \EOS [0]{\spacefactor3000\relax}%
\providecommand \BibitemShut  [1]{\csname bibitem#1\endcsname}%
\let\auto@bib@innerbib\@empty
\bibitem [{\citenamefont {Polchinski}(1999)}]{polchinskinotes}%
  \BibitemOpen
  \bibfield  {author} {\bibinfo {author} {\bibfnamefont {J.}~\bibnamefont {Polchinski}},\ }\href {https://arxiv.org/abs/hep-th/9210046} {\bibinfo {title} {Effective field theory and the fermi surface}} (\bibinfo {year} {1999}),\ \Eprint {https://arxiv.org/abs/hep-th/9210046} {arXiv:hep-th/9210046 [hep-th]} \BibitemShut {NoStop}%
\bibitem [{\citenamefont {{Landsman}}\ and\ \citenamefont {{van Weert}}(1987)}]{LandvWeert}%
  \BibitemOpen
  \bibfield  {author} {\bibinfo {author} {\bibfnamefont {N.~P.}\ \bibnamefont {{Landsman}}}\ and\ \bibinfo {author} {\bibfnamefont {C.~G.}\ \bibnamefont {{van Weert}}},\ }\bibfield  {title} {\bibinfo {title} {{Real- and imaginary-time field theory at finite temperature and density}},\ }\href {https://doi.org/10.1016/0370-1573(87)90121-9} {\bibfield  {journal} {\bibinfo  {journal} {physrep}\ }\textbf {\bibinfo {volume} {145}},\ \bibinfo {pages} {141} (\bibinfo {year} {1987})}\BibitemShut {NoStop}%
\bibitem [{\citenamefont {Das}(1997)}]{Das:ftft}%
  \BibitemOpen
  \bibfield  {author} {\bibinfo {author} {\bibfnamefont {A.~K.}\ \bibnamefont {Das}},\ }\href@noop {} {\emph {\bibinfo {title} {{Finite Temperature Field Theory}}}}\ (\bibinfo  {publisher} {World Scientific},\ \bibinfo {address} {New York},\ \bibinfo {year} {1997})\BibitemShut {NoStop}%
\bibitem [{\citenamefont {Calzetta}\ and\ \citenamefont {Hu}(2008)}]{calzetta2008nonequilibrium}%
  \BibitemOpen
  \bibfield  {author} {\bibinfo {author} {\bibfnamefont {E.}~\bibnamefont {Calzetta}}\ and\ \bibinfo {author} {\bibfnamefont {B.}~\bibnamefont {Hu}},\ }\href {https://books.google.si/books?id=BRJ7ryt2l1IC} {\emph {\bibinfo {title} {Nonequilibrium Quantum Field Theory}}},\ Cambridge Monographs on Mathematical Physics\ (\bibinfo  {publisher} {Cambridge University Press},\ \bibinfo {year} {2008})\BibitemShut {NoStop}%
\bibitem [{\citenamefont {Kamenev}(2011)}]{Kamenev_2011}%
  \BibitemOpen
  \bibfield  {author} {\bibinfo {author} {\bibfnamefont {A.}~\bibnamefont {Kamenev}},\ }\href@noop {} {\emph {\bibinfo {title} {Field Theory of Non-Equilibrium Systems}}}\ (\bibinfo  {publisher} {Cambridge University Press},\ \bibinfo {year} {2011})\BibitemShut {NoStop}%
\bibitem [{\citenamefont {Liu}\ and\ \citenamefont {Glorioso}(2018)}]{cgl_lectures}%
  \BibitemOpen
  \bibfield  {author} {\bibinfo {author} {\bibfnamefont {H.}~\bibnamefont {Liu}}\ and\ \bibinfo {author} {\bibfnamefont {P.}~\bibnamefont {Glorioso}},\ }\bibfield  {title} {\bibinfo {title} {{Lectures on non-equilibrium effective field theories and fluctuating hydrodynamics}},\ }\href {https://doi.org/10.22323/1.305.0008} {\bibfield  {journal} {\bibinfo  {journal} {PoS}\ }\textbf {\bibinfo {volume} {TASI2017}},\ \bibinfo {pages} {008} (\bibinfo {year} {2018})},\ \Eprint {https://arxiv.org/abs/1805.09331} {arXiv:1805.09331 [hep-th]} \BibitemShut {NoStop}%
\bibitem [{\citenamefont {Green}\ and\ \citenamefont {Sun}(2024)}]{Green:2024cmx}%
  \BibitemOpen
  \bibfield  {author} {\bibinfo {author} {\bibfnamefont {D.}~\bibnamefont {Green}}\ and\ \bibinfo {author} {\bibfnamefont {G.}~\bibnamefont {Sun}},\ }\bibfield  {title} {\bibinfo {title} {{Effective Field Theory and In-In Correlators}},\ }\Eprint {https://arxiv.org/abs/2412.02739} {arXiv:2412.02739 [hep-th]}  (\bibinfo {year} {2024})\BibitemShut {NoStop}%
\bibitem [{\citenamefont {Grozdanov}\ and\ \citenamefont {Polonyi}(2015)}]{sasojanos1}%
  \BibitemOpen
  \bibfield  {author} {\bibinfo {author} {\bibfnamefont {S.}~\bibnamefont {Grozdanov}}\ and\ \bibinfo {author} {\bibfnamefont {J.}~\bibnamefont {Polonyi}},\ }\bibfield  {title} {\bibinfo {title} {{Viscosity and dissipative hydrodynamics from effective field theory}},\ }\href {https://doi.org/10.1103/PhysRevD.91.105031} {\bibfield  {journal} {\bibinfo  {journal} {Phys. Rev. D}\ }\textbf {\bibinfo {volume} {91}},\ \bibinfo {pages} {105031} (\bibinfo {year} {2015})},\ \Eprint {https://arxiv.org/abs/1305.3670} {arXiv:1305.3670 [hep-th]} \BibitemShut {NoStop}%
\bibitem [{\citenamefont {Montenegro}\ and\ \citenamefont {Torrieri}(2016)}]{Montenegro:2016gjq}%
  \BibitemOpen
  \bibfield  {author} {\bibinfo {author} {\bibfnamefont {D.}~\bibnamefont {Montenegro}}\ and\ \bibinfo {author} {\bibfnamefont {G.}~\bibnamefont {Torrieri}},\ }\bibfield  {title} {\bibinfo {title} {{Lagrangian formulation of relativistic Israel-Stewart hydrodynamics}},\ }\href {https://doi.org/10.1103/PhysRevD.94.065042} {\bibfield  {journal} {\bibinfo  {journal} {Phys. Rev. D}\ }\textbf {\bibinfo {volume} {94}},\ \bibinfo {pages} {065042} (\bibinfo {year} {2016})},\ \Eprint {https://arxiv.org/abs/1604.05291} {arXiv:1604.05291 [hep-th]} \BibitemShut {NoStop}%
\bibitem [{\citenamefont {Crossley}\ \emph {et~al.}(2017)\citenamefont {Crossley}, \citenamefont {Glorioso},\ and\ \citenamefont {Liu}}]{Crossley_eft}%
  \BibitemOpen
  \bibfield  {author} {\bibinfo {author} {\bibfnamefont {M.}~\bibnamefont {Crossley}}, \bibinfo {author} {\bibfnamefont {P.}~\bibnamefont {Glorioso}},\ and\ \bibinfo {author} {\bibfnamefont {H.}~\bibnamefont {Liu}},\ }\bibfield  {title} {\bibinfo {title} {{Effective field theory of dissipative fluids}},\ }\href {https://doi.org/10.1007/JHEP09(2017)095} {\bibfield  {journal} {\bibinfo  {journal} {JHEP}\ }\textbf {\bibinfo {volume} {09}},\ \bibinfo {pages} {095}},\ \Eprint {https://arxiv.org/abs/1511.03646} {arXiv:1511.03646 [hep-th]} \BibitemShut {NoStop}%
\bibitem [{\citenamefont {Glorioso}\ and\ \citenamefont {Liu}(2016)}]{Glorioso:IIthermo}%
  \BibitemOpen
  \bibfield  {author} {\bibinfo {author} {\bibfnamefont {P.}~\bibnamefont {Glorioso}}\ and\ \bibinfo {author} {\bibfnamefont {H.}~\bibnamefont {Liu}},\ }\href@noop {} {\bibinfo {title} {{The second law of thermodynamics from symmetry and unitarity}}} (\bibinfo {year} {2016}),\ \Eprint {https://arxiv.org/abs/1612.07705} {arXiv:1612.07705 [hep-th]} \BibitemShut {NoStop}%
\bibitem [{\citenamefont {Glorioso}\ \emph {et~al.}(2017)\citenamefont {Glorioso}, \citenamefont {Crossley},\ and\ \citenamefont {Liu}}]{cgl_part2}%
  \BibitemOpen
  \bibfield  {author} {\bibinfo {author} {\bibfnamefont {P.}~\bibnamefont {Glorioso}}, \bibinfo {author} {\bibfnamefont {M.}~\bibnamefont {Crossley}},\ and\ \bibinfo {author} {\bibfnamefont {H.}~\bibnamefont {Liu}},\ }\bibfield  {title} {\bibinfo {title} {{Effective field theory of dissipative fluids (II): classical limit, dynamical KMS symmetry and entropy current}},\ }\href {https://doi.org/10.1007/JHEP09(2017)096} {\bibfield  {journal} {\bibinfo  {journal} {JHEP}\ }\textbf {\bibinfo {volume} {09}},\ \bibinfo {pages} {096}},\ \Eprint {https://arxiv.org/abs/1701.07817} {arXiv:1701.07817 [hep-th]} \BibitemShut {NoStop}%
\bibitem [{\citenamefont {Haehl}\ \emph {et~al.}(2014)\citenamefont {Haehl}, \citenamefont {Loganayagam},\ and\ \citenamefont {Rangamani}}]{HaehlRang_a}%
  \BibitemOpen
  \bibfield  {author} {\bibinfo {author} {\bibfnamefont {F.~M.}\ \bibnamefont {Haehl}}, \bibinfo {author} {\bibfnamefont {R.}~\bibnamefont {Loganayagam}},\ and\ \bibinfo {author} {\bibfnamefont {M.}~\bibnamefont {Rangamani}},\ }\bibfield  {title} {\bibinfo {title} {{Effective actions for anomalous hydrodynamics}},\ }\href {https://doi.org/10.1007/JHEP03(2014)034} {\bibfield  {journal} {\bibinfo  {journal} {JHEP}\ }\textbf {\bibinfo {volume} {03}},\ \bibinfo {pages} {034}},\ \Eprint {https://arxiv.org/abs/1312.0610} {arXiv:1312.0610 [hep-th]} \BibitemShut {NoStop}%
\bibitem [{\citenamefont {Haehl}\ \emph {et~al.}(2016{\natexlab{a}})\citenamefont {Haehl}, \citenamefont {Loganayagam},\ and\ \citenamefont {Rangamani}}]{HaehlRang_b}%
  \BibitemOpen
  \bibfield  {author} {\bibinfo {author} {\bibfnamefont {F.~M.}\ \bibnamefont {Haehl}}, \bibinfo {author} {\bibfnamefont {R.}~\bibnamefont {Loganayagam}},\ and\ \bibinfo {author} {\bibfnamefont {M.}~\bibnamefont {Rangamani}},\ }\bibfield  {title} {\bibinfo {title} {The fluid manifesto: emergent symmetries, hydrodynamics, and black holes},\ }\bibfield  {journal} {\bibinfo  {journal} {Journal of High Energy Physics}\ }\textbf {\bibinfo {volume} {2016}},\ \href {https://doi.org/10.1007/jhep01(2016)184} {10.1007/jhep01(2016)184} (\bibinfo {year} {2016}{\natexlab{a}})\BibitemShut {NoStop}%
\bibitem [{\citenamefont {Haehl}\ \emph {et~al.}(2016{\natexlab{b}})\citenamefont {Haehl}, \citenamefont {Loganayagam},\ and\ \citenamefont {Rangamani}}]{HaehlRang_sigma}%
  \BibitemOpen
  \bibfield  {author} {\bibinfo {author} {\bibfnamefont {F.~M.}\ \bibnamefont {Haehl}}, \bibinfo {author} {\bibfnamefont {R.}~\bibnamefont {Loganayagam}},\ and\ \bibinfo {author} {\bibfnamefont {M.}~\bibnamefont {Rangamani}},\ }\bibfield  {title} {\bibinfo {title} {{Topological sigma models \& dissipative hydrodynamics}},\ }\href {https://doi.org/10.1007/JHEP04(2016)039} {\bibfield  {journal} {\bibinfo  {journal} {JHEP}\ }\textbf {\bibinfo {volume} {04}},\ \bibinfo {pages} {039}},\ \Eprint {https://arxiv.org/abs/1511.07809} {arXiv:1511.07809 [hep-th]} \BibitemShut {NoStop}%
\bibitem [{\citenamefont {Chen-Lin}\ \emph {et~al.}(2019)\citenamefont {Chen-Lin}, \citenamefont {Delacr\'etaz},\ and\ \citenamefont {Hartnoll}}]{Chen-Lin:2018kfl}%
  \BibitemOpen
  \bibfield  {author} {\bibinfo {author} {\bibfnamefont {X.}~\bibnamefont {Chen-Lin}}, \bibinfo {author} {\bibfnamefont {L.~V.}\ \bibnamefont {Delacr\'etaz}},\ and\ \bibinfo {author} {\bibfnamefont {S.~A.}\ \bibnamefont {Hartnoll}},\ }\bibfield  {title} {\bibinfo {title} {{Theory of diffusive fluctuations}},\ }\href {https://doi.org/10.1103/PhysRevLett.122.091602} {\bibfield  {journal} {\bibinfo  {journal} {Phys. Rev. Lett.}\ }\textbf {\bibinfo {volume} {122}},\ \bibinfo {pages} {091602} (\bibinfo {year} {2019})},\ \Eprint {https://arxiv.org/abs/1811.12540} {arXiv:1811.12540 [hep-th]} \BibitemShut {NoStop}%
\bibitem [{\citenamefont {Jensen}\ \emph {et~al.}(2018{\natexlab{a}})\citenamefont {Jensen}, \citenamefont {Pinzani-Fokeeva},\ and\ \citenamefont {Yarom}}]{Jensen:2017kzi}%
  \BibitemOpen
  \bibfield  {author} {\bibinfo {author} {\bibfnamefont {K.}~\bibnamefont {Jensen}}, \bibinfo {author} {\bibfnamefont {N.}~\bibnamefont {Pinzani-Fokeeva}},\ and\ \bibinfo {author} {\bibfnamefont {A.}~\bibnamefont {Yarom}},\ }\bibfield  {title} {\bibinfo {title} {{Dissipative hydrodynamics in superspace}},\ }\href {https://doi.org/10.1007/JHEP09(2018)127} {\bibfield  {journal} {\bibinfo  {journal} {JHEP}\ }\textbf {\bibinfo {volume} {09}},\ \bibinfo {pages} {127}},\ \Eprint {https://arxiv.org/abs/1701.07436} {arXiv:1701.07436 [hep-th]} \BibitemShut {NoStop}%
\bibitem [{\citenamefont {Jensen}\ \emph {et~al.}(2018{\natexlab{b}})\citenamefont {Jensen}, \citenamefont {Marjieh}, \citenamefont {Pinzani-Fokeeva},\ and\ \citenamefont {Yarom}}]{Jensen:2018hse}%
  \BibitemOpen
  \bibfield  {author} {\bibinfo {author} {\bibfnamefont {K.}~\bibnamefont {Jensen}}, \bibinfo {author} {\bibfnamefont {R.}~\bibnamefont {Marjieh}}, \bibinfo {author} {\bibfnamefont {N.}~\bibnamefont {Pinzani-Fokeeva}},\ and\ \bibinfo {author} {\bibfnamefont {A.}~\bibnamefont {Yarom}},\ }\bibfield  {title} {\bibinfo {title} {{A panoply of Schwinger-Keldysh transport}},\ }\href {https://doi.org/10.21468/SciPostPhys.5.5.053} {\bibfield  {journal} {\bibinfo  {journal} {SciPost Phys.}\ }\textbf {\bibinfo {volume} {5}},\ \bibinfo {pages} {053} (\bibinfo {year} {2018}{\natexlab{b}})},\ \Eprint {https://arxiv.org/abs/1804.04654} {arXiv:1804.04654 [hep-th]} \BibitemShut {NoStop}%
\bibitem [{\citenamefont {Vardhan}\ \emph {et~al.}(2024{\natexlab{a}})\citenamefont {Vardhan}, \citenamefont {Grozdanov}, \citenamefont {Leutheusser},\ and\ \citenamefont {Liu}}]{Vardhan:short}%
  \BibitemOpen
  \bibfield  {author} {\bibinfo {author} {\bibfnamefont {S.}~\bibnamefont {Vardhan}}, \bibinfo {author} {\bibfnamefont {S.}~\bibnamefont {Grozdanov}}, \bibinfo {author} {\bibfnamefont {S.}~\bibnamefont {Leutheusser}},\ and\ \bibinfo {author} {\bibfnamefont {H.}~\bibnamefont {Liu}},\ }\bibfield  {title} {\bibinfo {title} {{Strong-field magnetohydrodynamics for neutron stars}},\ }\href {https://doi.org/10.1103/PhysRevResearch.6.L042050} {\bibfield  {journal} {\bibinfo  {journal} {Phys. Rev. Res.}\ }\textbf {\bibinfo {volume} {6}},\ \bibinfo {pages} {L042050} (\bibinfo {year} {2024}{\natexlab{a}})},\ \Eprint {https://arxiv.org/abs/2207.01636} {arXiv:2207.01636 [astro-ph.HE]} \BibitemShut {NoStop}%
\bibitem [{\citenamefont {Vardhan}\ \emph {et~al.}(2024{\natexlab{b}})\citenamefont {Vardhan}, \citenamefont {Grozdanov}, \citenamefont {Leutheusser},\ and\ \citenamefont {Liu}}]{Vardhan:long}%
  \BibitemOpen
  \bibfield  {author} {\bibinfo {author} {\bibfnamefont {S.}~\bibnamefont {Vardhan}}, \bibinfo {author} {\bibfnamefont {S.}~\bibnamefont {Grozdanov}}, \bibinfo {author} {\bibfnamefont {S.}~\bibnamefont {Leutheusser}},\ and\ \bibinfo {author} {\bibfnamefont {H.}~\bibnamefont {Liu}},\ }\href@noop {} {\bibinfo {title} {{Effective field theories of dissipative fluids with one-form symmetries}}} (\bibinfo {year} {2024}{\natexlab{b}}),\ \Eprint {https://arxiv.org/abs/2408.12868} {arXiv:2408.12868 [hep-th]} \BibitemShut {NoStop}%
\bibitem [{\citenamefont {Blake}\ \emph {et~al.}(2018)\citenamefont {Blake}, \citenamefont {Lee},\ and\ \citenamefont {Liu}}]{BlakeLiu1}%
  \BibitemOpen
  \bibfield  {author} {\bibinfo {author} {\bibfnamefont {M.}~\bibnamefont {Blake}}, \bibinfo {author} {\bibfnamefont {H.}~\bibnamefont {Lee}},\ and\ \bibinfo {author} {\bibfnamefont {H.}~\bibnamefont {Liu}},\ }\bibfield  {title} {\bibinfo {title} {{A quantum hydrodynamical description for scrambling and many-body chaos}},\ }\href {https://doi.org/10.1007/JHEP10(2018)127} {\bibfield  {journal} {\bibinfo  {journal} {JHEP}\ }\textbf {\bibinfo {volume} {10}},\ \bibinfo {pages} {127}},\ \Eprint {https://arxiv.org/abs/1801.00010} {arXiv:1801.00010 [hep-th]} \BibitemShut {NoStop}%
\bibitem [{\citenamefont {Blake}\ and\ \citenamefont {Liu}(2021)}]{BlakeLiu2}%
  \BibitemOpen
  \bibfield  {author} {\bibinfo {author} {\bibfnamefont {M.}~\bibnamefont {Blake}}\ and\ \bibinfo {author} {\bibfnamefont {H.}~\bibnamefont {Liu}},\ }\bibfield  {title} {\bibinfo {title} {{On systems of maximal quantum chaos}},\ }\href {https://doi.org/10.1007/JHEP05(2021)229} {\bibfield  {journal} {\bibinfo  {journal} {JHEP}\ }\textbf {\bibinfo {volume} {05}},\ \bibinfo {pages} {229}},\ \Eprint {https://arxiv.org/abs/2102.11294} {arXiv:2102.11294 [hep-th]} \BibitemShut {NoStop}%
\bibitem [{\citenamefont {Gao}\ and\ \citenamefont {Liu}(2023)}]{GaoLiu_nonmaximal}%
  \BibitemOpen
  \bibfield  {author} {\bibinfo {author} {\bibfnamefont {P.}~\bibnamefont {Gao}}\ and\ \bibinfo {author} {\bibfnamefont {H.}~\bibnamefont {Liu}},\ }\bibfield  {title} {\bibinfo {title} {{An effective field theory for non-maximal quantum chaos}},\ }\href {https://doi.org/10.1007/JHEP11(2023)076} {\bibfield  {journal} {\bibinfo  {journal} {JHEP}\ }\textbf {\bibinfo {volume} {11}},\ \bibinfo {pages} {076}},\ \Eprint {https://arxiv.org/abs/2301.05256} {arXiv:2301.05256 [hep-th]} \BibitemShut {NoStop}%
\bibitem [{\citenamefont {Winer}\ and\ \citenamefont {Swingle}(2022)}]{Winer:2020gdp}%
  \BibitemOpen
  \bibfield  {author} {\bibinfo {author} {\bibfnamefont {M.}~\bibnamefont {Winer}}\ and\ \bibinfo {author} {\bibfnamefont {B.}~\bibnamefont {Swingle}},\ }\bibfield  {title} {\bibinfo {title} {{Hydrodynamic Theory of the Connected Spectral form Factor}},\ }\href {https://doi.org/10.1103/PhysRevX.12.021009} {\bibfield  {journal} {\bibinfo  {journal} {Phys. Rev. X}\ }\textbf {\bibinfo {volume} {12}},\ \bibinfo {pages} {021009} (\bibinfo {year} {2022})},\ \Eprint {https://arxiv.org/abs/2012.01436} {arXiv:2012.01436 [cond-mat.stat-mech]} \BibitemShut {NoStop}%
\bibitem [{\citenamefont {Jain}\ and\ \citenamefont {Kovtun}(2022)}]{Jain:2020zhu}%
  \BibitemOpen
  \bibfield  {author} {\bibinfo {author} {\bibfnamefont {A.}~\bibnamefont {Jain}}\ and\ \bibinfo {author} {\bibfnamefont {P.}~\bibnamefont {Kovtun}},\ }\bibfield  {title} {\bibinfo {title} {{Late Time Correlations in Hydrodynamics: Beyond Constitutive Relations}},\ }\href {https://doi.org/10.1103/PhysRevLett.128.071601} {\bibfield  {journal} {\bibinfo  {journal} {Phys. Rev. Lett.}\ }\textbf {\bibinfo {volume} {128}},\ \bibinfo {pages} {071601} (\bibinfo {year} {2022})},\ \Eprint {https://arxiv.org/abs/2009.01356} {arXiv:2009.01356 [hep-th]} \BibitemShut {NoStop}%
\bibitem [{\citenamefont {Abbasi}\ and\ \citenamefont {Rischke}(2024)}]{Abbasi:2024pwz}%
  \BibitemOpen
  \bibfield  {author} {\bibinfo {author} {\bibfnamefont {N.}~\bibnamefont {Abbasi}}\ and\ \bibinfo {author} {\bibfnamefont {D.~H.}\ \bibnamefont {Rischke}},\ }\bibfield  {title} {\bibinfo {title} {{Three-point functions from a Schwinger-Keldysh effective action, resummed in derivatives}},\ }\Eprint {https://arxiv.org/abs/2410.07929} {arXiv:2410.07929 [hep-th]}  (\bibinfo {year} {2024})\BibitemShut {NoStop}%
\bibitem [{\citenamefont {Montenegro}\ \emph {et~al.}(2017)\citenamefont {Montenegro}, \citenamefont {Tinti},\ and\ \citenamefont {Torrieri}}]{Montenegro:2017rbu}%
  \BibitemOpen
  \bibfield  {author} {\bibinfo {author} {\bibfnamefont {D.}~\bibnamefont {Montenegro}}, \bibinfo {author} {\bibfnamefont {L.}~\bibnamefont {Tinti}},\ and\ \bibinfo {author} {\bibfnamefont {G.}~\bibnamefont {Torrieri}},\ }\bibfield  {title} {\bibinfo {title} {{Ideal relativistic fluid limit for a medium with polarization}},\ }\href {https://doi.org/10.1103/PhysRevD.96.056012} {\bibfield  {journal} {\bibinfo  {journal} {Phys. Rev. D}\ }\textbf {\bibinfo {volume} {96}},\ \bibinfo {pages} {056012} (\bibinfo {year} {2017})},\ \bibinfo {note} {[Addendum: Phys.Rev.D 96, 079901 (2017)]},\ \Eprint {https://arxiv.org/abs/1701.08263} {arXiv:1701.08263 [hep-th]} \BibitemShut {NoStop}%
\bibitem [{\citenamefont {Akyuz}\ \emph {et~al.}(2024)\citenamefont {Akyuz}, \citenamefont {Goon},\ and\ \citenamefont {Penco}}]{Akyuz:2023lsm}%
  \BibitemOpen
  \bibfield  {author} {\bibinfo {author} {\bibfnamefont {C.~O.}\ \bibnamefont {Akyuz}}, \bibinfo {author} {\bibfnamefont {G.}~\bibnamefont {Goon}},\ and\ \bibinfo {author} {\bibfnamefont {R.}~\bibnamefont {Penco}},\ }\bibfield  {title} {\bibinfo {title} {{The Schwinger-Keldysh coset construction}},\ }\href {https://doi.org/10.1007/JHEP06(2024)004} {\bibfield  {journal} {\bibinfo  {journal} {JHEP}\ }\textbf {\bibinfo {volume} {06}},\ \bibinfo {pages} {004}},\ \Eprint {https://arxiv.org/abs/2306.17232} {arXiv:2306.17232 [hep-th]} \BibitemShut {NoStop}%
\bibitem [{\citenamefont {Baggioli}\ \emph {et~al.}(2023)\citenamefont {Baggioli}, \citenamefont {Bu},\ and\ \citenamefont {Ziogas}}]{Baggioli:2023tlc}%
  \BibitemOpen
  \bibfield  {author} {\bibinfo {author} {\bibfnamefont {M.}~\bibnamefont {Baggioli}}, \bibinfo {author} {\bibfnamefont {Y.}~\bibnamefont {Bu}},\ and\ \bibinfo {author} {\bibfnamefont {V.}~\bibnamefont {Ziogas}},\ }\bibfield  {title} {\bibinfo {title} {{U(1) quasi-hydrodynamics: Schwinger-Keldysh effective field theory and holography}},\ }\href {https://doi.org/10.1007/JHEP09(2023)019} {\bibfield  {journal} {\bibinfo  {journal} {JHEP}\ }\textbf {\bibinfo {volume} {09}},\ \bibinfo {pages} {019}},\ \Eprint {https://arxiv.org/abs/2304.14173} {arXiv:2304.14173 [hep-th]} \BibitemShut {NoStop}%
\bibitem [{\citenamefont {Hongo}\ \emph {et~al.}(2024)\citenamefont {Hongo}, \citenamefont {Sogabe}, \citenamefont {Stephanov},\ and\ \citenamefont {Yee}}]{Hongo:2024brb}%
  \BibitemOpen
  \bibfield  {author} {\bibinfo {author} {\bibfnamefont {M.}~\bibnamefont {Hongo}}, \bibinfo {author} {\bibfnamefont {N.}~\bibnamefont {Sogabe}}, \bibinfo {author} {\bibfnamefont {M.~A.}\ \bibnamefont {Stephanov}},\ and\ \bibinfo {author} {\bibfnamefont {H.-U.}\ \bibnamefont {Yee}},\ }\bibfield  {title} {\bibinfo {title} {{Schwinger-Keldysh effective action for hydrodynamics with approximate symmetries}},\ }\Eprint {https://arxiv.org/abs/2411.08016} {arXiv:2411.08016 [hep-th]}  (\bibinfo {year} {2024})\BibitemShut {NoStop}%
\bibitem [{\citenamefont {Armas}\ \emph {et~al.}(2024)\citenamefont {Armas}, \citenamefont {Jain},\ and\ \citenamefont {Lier}}]{Armas:2024iuy}%
  \BibitemOpen
  \bibfield  {author} {\bibinfo {author} {\bibfnamefont {J.}~\bibnamefont {Armas}}, \bibinfo {author} {\bibfnamefont {A.}~\bibnamefont {Jain}},\ and\ \bibinfo {author} {\bibfnamefont {R.}~\bibnamefont {Lier}},\ }\bibfield  {title} {\bibinfo {title} {{Hydrodynamics of thermal active matter}},\ }\Eprint {https://arxiv.org/abs/2405.11023} {arXiv:2405.11023 [cond-mat.soft]}  (\bibinfo {year} {2024})\BibitemShut {NoStop}%
\bibitem [{\citenamefont {Jain}\ and\ \citenamefont {Kovtun}(2024)}]{Jain:2023obu}%
  \BibitemOpen
  \bibfield  {author} {\bibinfo {author} {\bibfnamefont {A.}~\bibnamefont {Jain}}\ and\ \bibinfo {author} {\bibfnamefont {P.}~\bibnamefont {Kovtun}},\ }\bibfield  {title} {\bibinfo {title} {{Schwinger-Keldysh effective field theory for stable and causal relativistic hydrodynamics}},\ }\href {https://doi.org/10.1007/JHEP01(2024)162} {\bibfield  {journal} {\bibinfo  {journal} {JHEP}\ }\textbf {\bibinfo {volume} {01}},\ \bibinfo {pages} {162}},\ \Eprint {https://arxiv.org/abs/2309.00511} {arXiv:2309.00511 [hep-th]} \BibitemShut {NoStop}%
\bibitem [{\citenamefont {Grozdanov}\ \emph {et~al.}(2024)\citenamefont {Grozdanov}, \citenamefont {Lemut}, \citenamefont {Pelai\v{c}},\ and\ \citenamefont {Soloviev}}]{Grozdanov:2024fle}%
  \BibitemOpen
  \bibfield  {author} {\bibinfo {author} {\bibfnamefont {S.}~\bibnamefont {Grozdanov}}, \bibinfo {author} {\bibfnamefont {T.}~\bibnamefont {Lemut}}, \bibinfo {author} {\bibfnamefont {J.}~\bibnamefont {Pelai\v{c}}},\ and\ \bibinfo {author} {\bibfnamefont {A.}~\bibnamefont {Soloviev}},\ }\bibfield  {title} {\bibinfo {title} {{Analytic structure of diffusive correlation functions}},\ }\href {https://doi.org/10.1103/PhysRevD.110.056053} {\bibfield  {journal} {\bibinfo  {journal} {Phys. Rev. D}\ }\textbf {\bibinfo {volume} {110}},\ \bibinfo {pages} {056053} (\bibinfo {year} {2024})},\ \Eprint {https://arxiv.org/abs/2407.13550} {arXiv:2407.13550 [hep-th]} \BibitemShut {NoStop}%
\bibitem [{\citenamefont {Nagy}\ and\ \citenamefont {Polonyi}(2022)}]{Nagy:2020bal}%
  \BibitemOpen
  \bibfield  {author} {\bibinfo {author} {\bibfnamefont {S.}~\bibnamefont {Nagy}}\ and\ \bibinfo {author} {\bibfnamefont {J.}~\bibnamefont {Polonyi}},\ }\bibfield  {title} {\bibinfo {title} {{Renormalizing Open Quantum Field Theories}},\ }\href {https://doi.org/10.3390/universe8020127} {\bibfield  {journal} {\bibinfo  {journal} {Universe}\ }\textbf {\bibinfo {volume} {8}},\ \bibinfo {pages} {127} (\bibinfo {year} {2022})},\ \Eprint {https://arxiv.org/abs/2012.13811} {arXiv:2012.13811 [hep-th]} \BibitemShut {NoStop}%
\bibitem [{\citenamefont {Kim}\ and\ \citenamefont {Ryu}(2023)}]{Kim:2023gaa}%
  \BibitemOpen
  \bibfield  {author} {\bibinfo {author} {\bibfnamefont {K.-S.}\ \bibnamefont {Kim}}\ and\ \bibinfo {author} {\bibfnamefont {S.}~\bibnamefont {Ryu}},\ }\bibfield  {title} {\bibinfo {title} {{Nonequilibrium thermodynamics perspectives for the monotonicity of the renormalization group flow}},\ }\href {https://doi.org/10.1103/PhysRevD.108.126022} {\bibfield  {journal} {\bibinfo  {journal} {Phys. Rev. D}\ }\textbf {\bibinfo {volume} {108}},\ \bibinfo {pages} {126022} (\bibinfo {year} {2023})},\ \Eprint {https://arxiv.org/abs/2310.15763} {arXiv:2310.15763 [hep-th]} \BibitemShut {NoStop}%
\bibitem [{\citenamefont {Jeon}(1993)}]{Jeon:spectral}%
  \BibitemOpen
  \bibfield  {author} {\bibinfo {author} {\bibfnamefont {S.}~\bibnamefont {Jeon}},\ }\bibfield  {title} {\bibinfo {title} {{Computing spectral densities in finite temperature field theory}},\ }\href {https://doi.org/10.1103/PhysRevD.47.4586} {\bibfield  {journal} {\bibinfo  {journal} {Phys. Rev. D}\ }\textbf {\bibinfo {volume} {47}},\ \bibinfo {pages} {4586} (\bibinfo {year} {1993})},\ \Eprint {https://arxiv.org/abs/hep-ph/9210227} {arXiv:hep-ph/9210227} \BibitemShut {NoStop}%
\bibitem [{\citenamefont {Jeon}(1995)}]{jeon:long}%
  \BibitemOpen
  \bibfield  {author} {\bibinfo {author} {\bibfnamefont {S.}~\bibnamefont {Jeon}},\ }\bibfield  {title} {\bibinfo {title} {{Hydrodynamic transport coefficients in relativistic scalar field theory}},\ }\href {https://doi.org/10.1103/PhysRevD.52.3591} {\bibfield  {journal} {\bibinfo  {journal} {Phys. Rev. D}\ }\textbf {\bibinfo {volume} {52}},\ \bibinfo {pages} {3591} (\bibinfo {year} {1995})}\BibitemShut {NoStop}%
\bibitem [{\citenamefont {Wang}\ and\ \citenamefont {Heinz}(2003)}]{Wang:2002nba}%
  \BibitemOpen
  \bibfield  {author} {\bibinfo {author} {\bibfnamefont {E.}~\bibnamefont {Wang}}\ and\ \bibinfo {author} {\bibfnamefont {U.~W.}\ \bibnamefont {Heinz}},\ }\bibfield  {title} {\bibinfo {title} {{Shear viscosity of hot scalar field theory in the real time formalism}},\ }\href {https://doi.org/10.1103/PhysRevD.67.025022} {\bibfield  {journal} {\bibinfo  {journal} {Phys. Rev. D}\ }\textbf {\bibinfo {volume} {67}},\ \bibinfo {pages} {025022} (\bibinfo {year} {2003})},\ \Eprint {https://arxiv.org/abs/hep-th/0201116} {arXiv:hep-th/0201116} \BibitemShut {NoStop}%
\bibitem [{\citenamefont {Arnold}\ \emph {et~al.}(2000)\citenamefont {Arnold}, \citenamefont {Moore},\ and\ \citenamefont {Yaffe}}]{AMYpaper}%
  \BibitemOpen
  \bibfield  {author} {\bibinfo {author} {\bibfnamefont {P.}~\bibnamefont {Arnold}}, \bibinfo {author} {\bibfnamefont {G.~D.}\ \bibnamefont {Moore}},\ and\ \bibinfo {author} {\bibfnamefont {L.~G.}\ \bibnamefont {Yaffe}},\ }\bibfield  {title} {\bibinfo {title} {{Transport coefficients in high temperature gauge theories (I): leading-log results}},\ }\href {https://doi.org/10.1088/1126-6708/2000/11/001} {\bibfield  {journal} {\bibinfo  {journal} {JHEP}\ }\textbf {\bibinfo {volume} {11}}\bibinfo  {number} { (2000)},\ \bibinfo {pages} {001}}\BibitemShut {NoStop}%
\bibitem [{\citenamefont {Polonyi}(2019)}]{Polonyi:2018ykh}%
  \BibitemOpen
\bibfield  {number} {  }\bibfield  {author} {\bibinfo {author} {\bibfnamefont {J.}~\bibnamefont {Polonyi}},\ }\bibfield  {title} {\bibinfo {title} {{Boost invariant regulator for field theories}},\ }\href {https://doi.org/10.1142/S0217751X19500179} {\bibfield  {journal} {\bibinfo  {journal} {Int. J. Mod. Phys. A}\ }\textbf {\bibinfo {volume} {34}},\ \bibinfo {pages} {1950017} (\bibinfo {year} {2019})},\ \Eprint {https://arxiv.org/abs/1803.09292} {arXiv:1803.09292 [hep-th]} \BibitemShut {NoStop}%
\bibitem [{\citenamefont {Steib}\ \emph {et~al.}(2021)\citenamefont {Steib}, \citenamefont {Nagy},\ and\ \citenamefont {Polonyi}}]{Steib:2019xrv}%
  \BibitemOpen
  \bibfield  {author} {\bibinfo {author} {\bibfnamefont {I.}~\bibnamefont {Steib}}, \bibinfo {author} {\bibfnamefont {S.}~\bibnamefont {Nagy}},\ and\ \bibinfo {author} {\bibfnamefont {J.}~\bibnamefont {Polonyi}},\ }\bibfield  {title} {\bibinfo {title} {{Renormalization in Minkowski space\textendash{}time}},\ }\href {https://doi.org/10.1142/S0217751X21500317} {\bibfield  {journal} {\bibinfo  {journal} {Int. J. Mod. Phys. A}\ }\textbf {\bibinfo {volume} {36}},\ \bibinfo {pages} {2150031} (\bibinfo {year} {2021})},\ \Eprint {https://arxiv.org/abs/1908.11311} {arXiv:1908.11311 [hep-th]} \BibitemShut {NoStop}%
\bibitem [{\citenamefont {Ojima}(1986)}]{Ojima:nice}%
  \BibitemOpen
  \bibfield  {author} {\bibinfo {author} {\bibfnamefont {I.}~\bibnamefont {Ojima}},\ }\bibfield  {title} {\bibinfo {title} {{Lorentz Invariance Versus Temperature in {QFT}}},\ }\href {https://doi.org/10.1007/BF00417467} {\bibfield  {journal} {\bibinfo  {journal} {Lett. Math. Phys.}\ }\textbf {\bibinfo {volume} {11}},\ \bibinfo {pages} {73} (\bibinfo {year} {1986})}\BibitemShut {NoStop}%
\bibitem [{\citenamefont {Elmfors}\ and\ \citenamefont {Kobes}(1995)}]{ElmforsKobes_thbeta}%
  \BibitemOpen
  \bibfield  {author} {\bibinfo {author} {\bibfnamefont {P.}~\bibnamefont {Elmfors}}\ and\ \bibinfo {author} {\bibfnamefont {R.}~\bibnamefont {Kobes}},\ }\bibfield  {title} {\bibinfo {title} {{The Thermal beta function in Yang-Mills theory}},\ }\href {https://doi.org/10.1103/PhysRevD.51.774} {\bibfield  {journal} {\bibinfo  {journal} {Phys. Rev. D}\ }\textbf {\bibinfo {volume} {51}},\ \bibinfo {pages} {774} (\bibinfo {year} {1995})},\ \Eprint {https://arxiv.org/abs/hep-ph/9408254} {arXiv:hep-ph/9408254} \BibitemShut {NoStop}%
\bibitem [{\citenamefont {Niemi}\ and\ \citenamefont {Semenoff}(1984)}]{NiemiSemenoff_nuclb}%
  \BibitemOpen
  \bibfield  {author} {\bibinfo {author} {\bibfnamefont {A.~J.}\ \bibnamefont {Niemi}}\ and\ \bibinfo {author} {\bibfnamefont {G.~W.}\ \bibnamefont {Semenoff}},\ }\bibfield  {title} {\bibinfo {title} {{Thermodynamic Calculations in Relativistic Finite Temperature Quantum Field Theories}},\ }\href {https://doi.org/10.1016/0550-3213(84)90123-8} {\bibfield  {journal} {\bibinfo  {journal} {Nucl. Phys. B}\ }\textbf {\bibinfo {volume} {230}},\ \bibinfo {pages} {181} (\bibinfo {year} {1984})}\BibitemShut {NoStop}%
\bibitem [{\citenamefont {Stanford}(2016)}]{Stanford_wkchaos}%
  \BibitemOpen
  \bibfield  {author} {\bibinfo {author} {\bibfnamefont {D.}~\bibnamefont {Stanford}},\ }\bibfield  {title} {\bibinfo {title} {Many-body chaos at weak coupling},\ }\bibfield  {journal} {\bibinfo  {journal} {Journal of High Energy Physics}\ }\textbf {\bibinfo {volume} {2016}},\ \href {https://doi.org/10.1007/jhep10(2016)009} {10.1007/jhep10(2016)009} (\bibinfo {year} {2016})\BibitemShut {NoStop}%
\bibitem [{\citenamefont {Dolan}\ and\ \citenamefont {Jackiw}(1974)}]{DolanJackiw}%
  \BibitemOpen
  \bibfield  {author} {\bibinfo {author} {\bibfnamefont {L.}~\bibnamefont {Dolan}}\ and\ \bibinfo {author} {\bibfnamefont {R.}~\bibnamefont {Jackiw}},\ }\bibfield  {title} {\bibinfo {title} {{Symmetry Behavior at Finite Temperature}},\ }\href {https://doi.org/10.1103/PhysRevD.9.3320} {\bibfield  {journal} {\bibinfo  {journal} {Phys. Rev. D}\ }\textbf {\bibinfo {volume} {9}},\ \bibinfo {pages} {3320} (\bibinfo {year} {1974})}\BibitemShut {NoStop}%
\bibitem [{\citenamefont {Polonyi}(2003)}]{Polonyi:2001se}%
  \BibitemOpen
  \bibfield  {author} {\bibinfo {author} {\bibfnamefont {J.}~\bibnamefont {Polonyi}},\ }\bibfield  {title} {\bibinfo {title} {{Lectures on the functional renormalization group method}},\ }\href {https://doi.org/10.2478/BF02475552} {\bibfield  {journal} {\bibinfo  {journal} {Central Eur. J. Phys.}\ }\textbf {\bibinfo {volume} {1}},\ \bibinfo {pages} {1} (\bibinfo {year} {2003})},\ \Eprint {https://arxiv.org/abs/hep-th/0110026} {arXiv:hep-th/0110026} \BibitemShut {NoStop}%
\bibitem [{\citenamefont {Zamolodchikov}(1986)}]{zamolodchikov:Ctheo}%
  \BibitemOpen
  \bibfield  {author} {\bibinfo {author} {\bibfnamefont {A.~B.}\ \bibnamefont {Zamolodchikov}},\ }\bibfield  {title} {\bibinfo {title} {Irreversibility of the flux of the renormalization group in a 2d field theory},\ }\href@noop {} {\bibfield  {journal} {\bibinfo  {journal} {JETP lett}\ }\textbf {\bibinfo {volume} {43}},\ \bibinfo {pages} {730} (\bibinfo {year} {1986})}\BibitemShut {NoStop}%
\bibitem [{\citenamefont {Jepsen}\ \emph {et~al.}(2021)\citenamefont {Jepsen}, \citenamefont {Klebanov},\ and\ \citenamefont {Popov}}]{JKP_rg1}%
  \BibitemOpen
  \bibfield  {author} {\bibinfo {author} {\bibfnamefont {C.~B.}\ \bibnamefont {Jepsen}}, \bibinfo {author} {\bibfnamefont {I.~R.}\ \bibnamefont {Klebanov}},\ and\ \bibinfo {author} {\bibfnamefont {F.~K.}\ \bibnamefont {Popov}},\ }\bibfield  {title} {\bibinfo {title} {{RG limit cycles and unconventional fixed points in perturbative QFT}},\ }\href {https://doi.org/10.1103/PhysRevD.103.046015} {\bibfield  {journal} {\bibinfo  {journal} {Phys. Rev. D}\ }\textbf {\bibinfo {volume} {103}},\ \bibinfo {pages} {046015} (\bibinfo {year} {2021})},\ \Eprint {https://arxiv.org/abs/2010.15133} {arXiv:2010.15133 [hep-th]} \BibitemShut {NoStop}%
\bibitem [{\citenamefont {Jepsen}\ and\ \citenamefont {Popov}(2021)}]{JP_rg2}%
  \BibitemOpen
  \bibfield  {author} {\bibinfo {author} {\bibfnamefont {C.~B.}\ \bibnamefont {Jepsen}}\ and\ \bibinfo {author} {\bibfnamefont {F.~K.}\ \bibnamefont {Popov}},\ }\bibfield  {title} {\bibinfo {title} {{Homoclinic Renormalization Group Flows, or When Relevant Operators Become Irrelevant}},\ }\href {https://doi.org/10.1103/PhysRevLett.127.141602} {\bibfield  {journal} {\bibinfo  {journal} {Phys. Rev. Lett.}\ }\textbf {\bibinfo {volume} {127}},\ \bibinfo {pages} {141602} (\bibinfo {year} {2021})},\ \Eprint {https://arxiv.org/abs/2105.01625} {arXiv:2105.01625 [hep-th]} \BibitemShut {NoStop}%
\bibitem [{\citenamefont {Nauenberg}(1975)}]{Nauenberg_1975}%
  \BibitemOpen
  \bibfield  {author} {\bibinfo {author} {\bibfnamefont {M.}~\bibnamefont {Nauenberg}},\ }\bibfield  {title} {\bibinfo {title} {Scaling representation for critical phenomena},\ }\href {https://doi.org/10.1088/0305-4470/8/6/011} {\bibfield  {journal} {\bibinfo  {journal} {Journal of Physics A: Mathematical and General}\ }\textbf {\bibinfo {volume} {8}},\ \bibinfo {pages} {925} (\bibinfo {year} {1975})}\BibitemShut {NoStop}%
\bibitem [{\citenamefont {Sornette}\ and\ \citenamefont {Sammis}(1995)}]{sornette:jpa-00247086}%
  \BibitemOpen
  \bibfield  {author} {\bibinfo {author} {\bibfnamefont {D.}~\bibnamefont {Sornette}}\ and\ \bibinfo {author} {\bibfnamefont {C.}~\bibnamefont {Sammis}},\ }\bibfield  {title} {\bibinfo {title} {{Complex Critical Exponents from Renormalization Group Theory of Earthquakes: Implications for Earthquake Predictions}},\ }\href {https://doi.org/10.1051/jp1:1995154} {\bibfield  {journal} {\bibinfo  {journal} {{Journal de Physique I}}\ }\textbf {\bibinfo {volume} {5}},\ \bibinfo {pages} {607} (\bibinfo {year} {1995})}\BibitemShut {NoStop}%
\bibitem [{\citenamefont {Jana}\ \emph {et~al.}(2020)\citenamefont {Jana}, \citenamefont {Loganayagam},\ and\ \citenamefont {Rangamani}}]{Jana:2020vyx}%
  \BibitemOpen
  \bibfield  {author} {\bibinfo {author} {\bibfnamefont {C.}~\bibnamefont {Jana}}, \bibinfo {author} {\bibfnamefont {R.}~\bibnamefont {Loganayagam}},\ and\ \bibinfo {author} {\bibfnamefont {M.}~\bibnamefont {Rangamani}},\ }\bibfield  {title} {\bibinfo {title} {{Open quantum systems and Schwinger-Keldysh holograms}},\ }\href {https://doi.org/10.1007/JHEP07(2020)242} {\bibfield  {journal} {\bibinfo  {journal} {JHEP}\ }\textbf {\bibinfo {volume} {07}},\ \bibinfo {pages} {242}},\ \Eprint {https://arxiv.org/abs/2004.02888} {arXiv:2004.02888 [hep-th]} \BibitemShut {NoStop}%
\bibitem [{\citenamefont {Loganayagam}\ \emph {et~al.}(2023)\citenamefont {Loganayagam}, \citenamefont {Rangamani},\ and\ \citenamefont {Virrueta}}]{Loganayagam:2022zmq}%
  \BibitemOpen
  \bibfield  {author} {\bibinfo {author} {\bibfnamefont {R.}~\bibnamefont {Loganayagam}}, \bibinfo {author} {\bibfnamefont {M.}~\bibnamefont {Rangamani}},\ and\ \bibinfo {author} {\bibfnamefont {J.}~\bibnamefont {Virrueta}},\ }\bibfield  {title} {\bibinfo {title} {{Holographic open quantum systems: toy models and analytic properties of thermal correlators}},\ }\href {https://doi.org/10.1007/JHEP03(2023)153} {\bibfield  {journal} {\bibinfo  {journal} {JHEP}\ }\textbf {\bibinfo {volume} {03}},\ \bibinfo {pages} {153}},\ \Eprint {https://arxiv.org/abs/2211.07683} {arXiv:2211.07683 [hep-th]} \BibitemShut {NoStop}%
\bibitem [{\citenamefont {Eden}\ \emph {et~al.}(1966)\citenamefont {Eden}, \citenamefont {Landshoff}, \citenamefont {Olive},\ and\ \citenamefont {Polkinghorne}}]{Eden_anSmatrix}%
  \BibitemOpen
  \bibfield  {author} {\bibinfo {author} {\bibfnamefont {R.~J.}\ \bibnamefont {Eden}}, \bibinfo {author} {\bibfnamefont {P.~V.}\ \bibnamefont {Landshoff}}, \bibinfo {author} {\bibfnamefont {D.~I.}\ \bibnamefont {Olive}},\ and\ \bibinfo {author} {\bibfnamefont {J.~C.}\ \bibnamefont {Polkinghorne}},\ }\href@noop {} {\emph {\bibinfo {title} {{The analytic S-matrix}}}}\ (\bibinfo  {publisher} {Cambridge Univ. Press},\ \bibinfo {address} {Cambridge},\ \bibinfo {year} {1966})\BibitemShut {NoStop}%
\end{thebibliography}%

\end{document}